\documentclass[onecolumn,aps,prx,floatfix,preprintnumbers,superscriptaddress]{revtex4}
\usepackage{graphicx}
\usepackage{psfig}
\usepackage{color}
\usepackage{appendix}
\usepackage{booktabs}
\usepackage[colorlinks=true, linkcolor=blue, citecolor=blue, urlcolor=blue]{hyperref}
\usepackage{verbatim}
\usepackage[normalem]{ulem}  

\newcommand{ \be }{\begin{equation}}
\newcommand{ \ee }{\end{equation}}
\newcommand{ \bea }{\begin{eqnarray}}
\newcommand{ \eea }{\end{eqnarray}}

\begin{document}
\preprint{RIKEN-iTHEMS-Report-26}
\title{Ultra-Peripheral Collisions as a Nuclear-Structure Interferometer with Interpretable Multitask Deep Learning}
    \author{Jing-Zong Zhang}
	\affiliation{Key Laboratory of Nuclear Physics and Ion-beam Application (MOE), Institute of Modern Physics, Fudan University, Shanghai 200433, China}
	\affiliation{Shanghai Research Center for Theoretical Nuclear Physics, NSFC and Fudan University, Shanghai $200438$, China}

    \author{Wang-Mei Zha}
    \email[]{first@ustc.edu.cn} 
    \affiliation{Department of Modern Physics, University of Science and Technology of China, Hefei 230026, China}

    \author{Lingxiao Wang}
    \email[]{lingxiao.wang@riken.jp}
    \affiliation{RIKEN Interdisciplinary Theoretical and Mathematical Sciences (iTHEMS), Wako, Saitama 351-0198, Japan}
    \affiliation{Institute for Physics of Intelligence, The University of Tokyo, Hongo,  Tokyo 113-0033, Japan}
    
	\author{Guo-Liang Ma}
	\email[]{glma@fudan.edu.cn}
	\affiliation{Key Laboratory of Nuclear Physics and Ion-beam Application (MOE), Institute of Modern Physics, Fudan University, Shanghai 200433, China}
	\affiliation{Shanghai Research Center for Theoretical Nuclear Physics, NSFC and Fudan University, Shanghai $200438$, China}
\begin{abstract}
Precise knowledge of nuclear structure is essential across fundamental physics, yet probing these structures is notoriously difficult. To address this challenge, ultra-peripheral collisions (UPCs) provide a femtoscopic tomography for imaging the atomic nucleus. UPCs offer a pristine electromagnetic pathway: coherent vector-meson photoproduction generates patterns of diffraction and two-source interference that directly encode the nuclear spatial density. Turning these patterns into quantitative constraints is, however, a challenging inverse problem, complicated by correlated sensitivities to deformation and neutron skin, phase smearing, and experimental backgrounds. Here we introduce an interpretable Multitask deep-learning framework that maps transverse momentum distributions to multiple nuclear-structure indicators simultaneously and identifies the kinematic regions driving each inference. We demonstrate the approach with coherent $J/\psi$ photoproduction in $^{96}_{40}\text{Zr} + ^{96}_{40}\text{Zr}$ collisions, showing that the learned features separate diffraction-dominated and interference-dominated information and provide analysis-ready observables for future high-luminosity data.
\end{abstract}
\maketitle


\section{INTRODUCTION}

Coherent imaging through interference and diffraction underpins some of the most powerful structural probes in modern physics, from x-ray ptychography and coherent diffractive imaging at angstrom scales~\cite{nazirkar2024coherent,ossig2024novel,miao1999extending,thibault2008high} to 4D-STEM mapping of nanoscale strain fields~\cite{ophus2019four,ozdol2015strain}. At their core, these techniques all solve the same inverse problem: recovering the real-space geometry of objects, e.g., nanocrystals~\cite{yau2017bragg,robinson2009coherent,chen2013three} and biological macromolecules~\cite{rodriguez2015three}, from intensity modulations imprinted in reciprocal space~\cite{yau2017bragg}. Can this framework be extended to the femtometer scale to image the atomic nucleus?

The spatial structure of the nuclear ground state is governed by a subtle interplay between two fundamental yet distinct sectors of nuclear physics. Collective shape deformation, characterized by parameters such as the quadrupole deformation $\beta_2$, encodes the interplay of shell structure and residual correlations that drives collective motion and low-energy spectroscopy~\cite{hamamoto2011shape,cline1985nuclear}. The radial distribution of protons and neutrons, quantified by the neutron-skin thickness $\Delta r_{np}$, constrains the density dependence of the symmetry energy and the equation of state governing neutron-star structure~\cite{ding2024neutron,fattoyev2012neutron,lattimer2004physics,roca2011neutron}. Simultaneously constraining both sectors in a single measurement is of central importance, as it would bridge low-energy nuclear structure with astrophysical observables. Yet classical probes such as elastic electron scattering and spectroscopy face a fundamental structural entanglement problem: signatures of intrinsic deformation and neutron skin are inextricably coupled in many observables~\cite{patterson2003empirical}, precluding a model-independent simultaneous extraction. To overcome this impasse, recent advances have turned to relativistic heavy-ion collisions as a snapshot camera for nuclear geometry~\cite{star2024imaging}. Among these, ultra-peripheral collisions (UPCs) provide a particularly direct route to coherent interference tomography of nuclear structure~\cite{ragoni2024overview}.

\begin{figure*}[htb]
\includegraphics[width=1\textwidth]{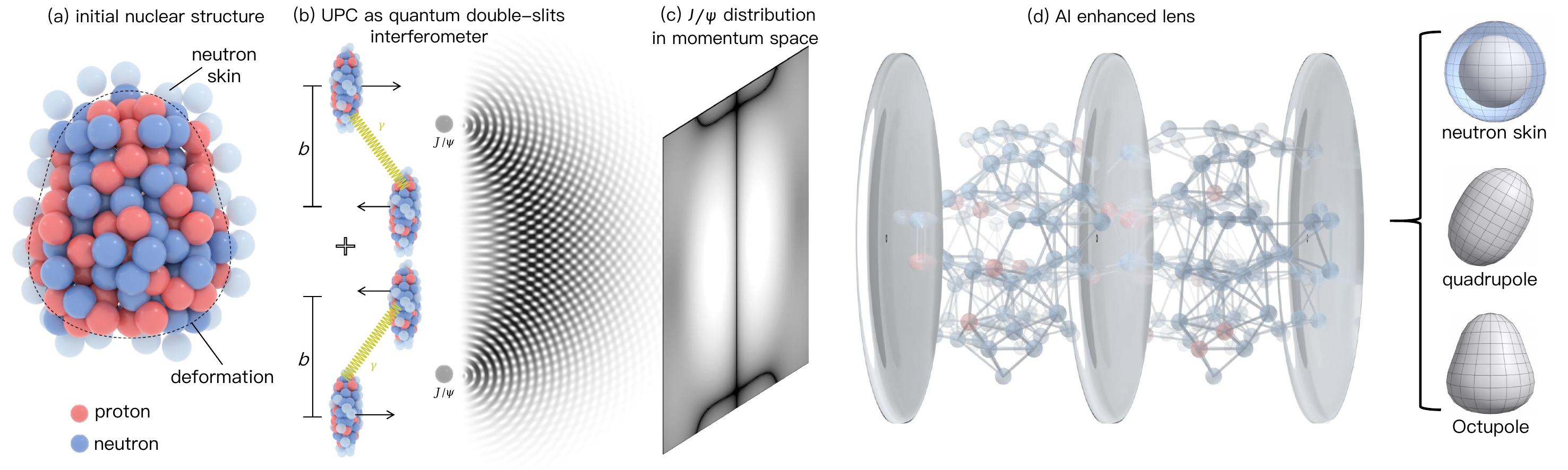}
\caption{\footnotesize(Color online) Nuclear interference tomography in ultra-peripheral collisions (UPCs).
(a) Initial nuclear configurations with neutron skin and shape deformations.
(b) The UPC system acting as a femtoscopic quantum interferometer with impact parameter $b$.
(c) The resulting two-dimensional transverse momentum ($p_\perp$) distribution of coherent $J/\psi$.
(d) The deep-learning framework ``AI-enhanced lens'') employed to reconstruct the initial nuclear-structure parameters from the final-state momentum image (e.g., neutron skin, quadrupole, and octupole).
} \label{fig:workflow}
\end{figure*}

As illustrated in Fig.~\ref{fig:workflow}, a UPC event constitutes a femtometer-scale quantum interferometer. The boosted electromagnetic field of one nucleus acts as the coherent laser source [Fig.~\ref{fig:workflow}(a)]; the two colliding nuclei, separated by an impact parameter $b$, function as a double-slit aperture [Fig.~\ref{fig:workflow}(b)]; and the coherent photoproduction of a vector meson (e.g., $J/\psi$, $\rho$) generates a momentum-space distribution that mirrors classical interference fringes modulated by a single-slit diffraction envelope [Fig.~\ref{fig:workflow}(c)]~\cite{shen2024exploring,luo2023effect,zha2019double}. Unlike hadronic collisions, this process is electromagnetic and exclusive, projecting the nuclear density profile directly onto the transverse momentum ($p_\perp$) distribution of the produced mesons~\cite{lin2025nuclear,abelev2009observation}. This quantum interference has been experimentally established in $\rho^0$ photoproduction~\cite{abelev2009observation}, and its interpretation as a Fermi-scale double-slit scenario has motivated a broad program of coherence- and interference-based nuclear imaging~\cite{zha2019double,ma2023new}.

Formally, the coherent photoproduction probability is proportionally related to the Fourier transform of the photon-nucleus scattering amplitudes $A_1$ and $A_2$, evaluated at positions displaced by the slit separation $b$:
\begin{equation}\label{Eq:Pr_Intro}
\frac{d^2 P}{d^2 \mathbf{p}_\perp} \propto \left| \mathcal{F}\left\{ A_1(\mathbf{r_\perp} - \mathbf{b}/2) - A_2(\mathbf{r_\perp} + \mathbf{b}/2) \right\} \right|^2.
\end{equation}
A qualitative analysis within the Glauber approximation reveals how different aspects of nuclear structure leave distinct imprints. In the limit of identical spherical nuclei, the probability factorizes into a structure-dependent diffraction term and a geometry-dependent interference term:
\begin{equation}\label{Eq:skin}
\frac{d^2P}{d^2\mathbf{p}_\perp} \propto \left| \mathcal{F}[A(\mathbf{r_\perp})] \right|^2 \sin^2\!\left( \frac{\mathbf{p}_\perp \cdot \mathbf{b}}{2} \right).
\end{equation}
Since the amplitude $A(\mathbf{r_\perp})$ depends on the nuclear density distribution, variations in the neutron-skin thickness directly modify its radial profile, thereby modulating the intensity of the diffraction bright spots. Conversely, nuclear deformation breaks the symmetry ($A_1 \neq A_2$) due to random spatial orientations of the two nuclei, introducing an event-by-event phase shift $\Delta\phi$ in the interference pattern:
\begin{equation}\label{Eq:def}
\frac{d^2P}{d^2\mathbf{p}_\perp} \propto F_1^2 + F_2^2 - 2|F_1 F_2| \cos\!\left( \mathbf{p}_\perp \cdot \mathbf{b} + \Delta\phi \right),
\end{equation}
where $F_i = \mathcal{F}[A_i(\mathbf{r_\perp})]$ and $\Delta\phi$ is the relative phase difference between $F_1$ and $F_2$. Therefore, the neutron skin governs the envelope of the diffraction peaks, while deformation controls the depth and position of the interference minima. In principle, these provide complementary and separable handles on the two structural sectors.

In practice, however, extracting both quantities simultaneously from the measured $p_T$ spectrum presents a challenging inverse problem~\cite{aarts2025physics}. At the \emph{fundamental} level, the difficulty stems from holographic degeneracy: in nuclear collisions, the nuclei are randomly oriented in each event, and experiments access only the ensemble average over these orientations. The stochastic fluctuation of $\Delta\phi$ does not simply shift the fringes but smears them, effectively filling in the interference minima and erasing the clean separation between diffraction and interference anticipated by Eqs.~(\ref{Eq:skin})--(\ref{Eq:def}). For nuclei of interest such as $^{96}$Zr, where significant deformation and neutron skin coexist~\cite{abdallah2022search,xu2021determine}, the two effects produce correlated distortions of the spectrum that defy disentanglement by simple inspection.

These intrinsic difficulties are further compounded at the experimental level. Finite detector resolution and soft transverse kicks smear the interference phases; incoherent photoproduction and other backgrounds contaminate the coherent signal; and experimental selections such as neutron tagging reshape the effective impact-parameter distribution~\cite{jia2023separating}. As a result, conventional fit-based extractions from a single spectrum suffer from parameter degeneracies and fragile uncertainty propagation when multiple geometric degrees of freedom are varied simultaneously. No simple analytical inversion exists to map the noisy, orientation-averaged interference pattern back to the intrinsic nuclear geometry.

Machine learning (ML) has been increasingly applied in high-energy nuclear physics~\cite{Boehnlein:2021eym,He:2023zin,zhou2024exploring}, with successful applications ranging from constraining the equation of state~\cite{aarts2025physics} and identifying phase transitions~\cite{Ma:2023zfj} to jet tomography~\cite{Du:2021pqa,Larkoski:2024uoc}. In the context of nuclear structure extraction, however, most existing approaches rely on Bayesian inference~\cite{Utama:2015hva,Cheng:2023ucp} built upon pre-defined, human-engineered observables as summary statistics. While robust for bulk properties, this paradigm performs a lossy compression of the data. Integrating over the momentum-space distribution to obtain scalar quantities washes out or inextricably entangles the subtle localized spectral correlations that distinguish deformation from neutron-skin effects. To overcome these limitations, we employ a deep-learning framework, conceptualized as an ``AI-enhanced lens'' [Fig.~\ref{fig:workflow}(d)], that operates directly on the full $J/\psi$ transverse momentum spectrum. By learning a nonlinear inverse mapping from complete spectral shapes, the network preserves the localized features encoding the interplay of deformation and neutron skin, and is trained to be robust against the orientation averaging, detector smearing, and background contamination inherent in realistic UPC data. In this work, we demonstrate that this approach can simultaneously extract the quadrupole deformation $\beta_2$ and the neutron-skin thickness of $^{96}$Zr from coherent $J/\psi$ photoproduction in UPCs, breaking the degeneracy that has limited conventional analyses.

To resolve this impasse, Sec.~\ref{sec:model} introduces a physics-driven Multitask deep-learning framework designed to disentangle these competing geometric signatures. Here, we model coherent vector-meson photoproduction within a vector-meson-dominance and Glauber-based description across ensembles of nuclear configurations spanning realistic shape and skin variations, and we couple these calculations to an interpretable Multitask deep-learning approach~\cite{Boehnlein:2021eym} that infers multiple nuclear-structure indicators simultaneously while diagnosing which kinematic regions provide the discriminating power for each inference. Our central idea, explored in Sec.~\ref{sec:interpre}, is to treat the two-dimensional transverse momentum distribution of coherent vector mesons as an interference image, and to train a shared feature extractor with task-specific heads that target shape regression and neutron-skin classification within a unified model. Section~\ref{sec:obser} then connects the learned features back to physics by identifying robust, analysis-ready observables that isolate diffraction-dominated bright-spot information from interference-dominated fringe information, providing a transparent bridge between machine-learning inference and experimentally accessible kinematic windows. Using coherent $J/\psi$ photoproduction in $^{96}\mathrm{Zr}$+$^{96}\mathrm{Zr}$ collisions at $\sqrt{s_{NN}}=200$~GeV, we show that the learned representation yields a physically meaningful separation between diffraction-driven sensitivity to the neutron skin and interference-driven sensitivity to deformation. Furthermore, we show that these metrics can map to experimental observables such as anisotropic flow coefficients $v_n$ and low-$p_T$ yield ratios in specific kinematic windows, enabling quantitative constraints from future high-luminosity UPC data at the Relativistic Heavy Ion Collider (RHIC) and the Large Hadron Collider (LHC). Finally, a summary and outlook are provided in Sec.~\ref{sec:conclu}.


\section{PHYSICS MODELING AND MULTITASK DEEP-LEARNING FRAMEWORK}\label{sec:model}

\subsection{Physics modeling}

$J/\psi$ photoproduction in $^{96}_{40}\text{Zr}$+$^{96}_{40}\text{Zr}$ ultra-peripheral collisions is simulated using the Glauber vector-meson-dominance (Glauber-VMD) framework
established in Refs.~\cite{shen2024exploring,luo2023effect,zha2019double}.
The spatial distribution of nucleons inside the $^{96}_{40}\text{Zr}$ nucleus is modeled
with a deformed Woods--Saxon profile:

\begin{equation}
\rho_A(r,\theta,\phi) = \frac{\rho^0}{1+\exp[(r-R(\theta,\phi))/a]},
\end{equation}
where the radius parameter $R(\theta,\phi)$ incorporates deformations:

\begin{equation}
R(\theta,\phi) = R_0\left [ 1+\beta_2 Y_{2,0}(\theta , \phi) +\beta_3 Y_{3,0}(\theta , \phi)   \right ].
\end{equation}
Here, $\rho^0$ is the normal nuclear density, 
$R_0$ is the nuclear radius, 
$a$ is the surface diffuseness parameter, 
and $\beta_2$ and $\beta_3$ represent the quadrupole and octupole deformation parameters, respectively.
To model the neutron-skin effect, we distinguish between the proton and neutron density distributions. Specifically, a skin-type neutron skin is implemented by assigning a larger radius to neutrons than protons ($R_{0}^n>R_0^p$) while keeping the diffuseness $a$ identical. Conversely, a halo-type structure is characterized by a larger surface diffuseness for neutrons ($a_n>a_p$) with similar radii.

The coherent photoproduction amplitude is constructed by convolving the equivalent photon flux~\cite{krauss1997photon} with the photon-nucleus scattering amplitude, incorporating nuclear shadowing and coherence-length effects~\cite{miller2007glauber,bauer1978hadronic}. The two-dimensional transverse momentum distribution is then obtained via a Fourier transformation of the coordinate space amplitudes from the two interference terms ($A_1$ and $A_2$):
\begin{equation}\label{Eq:Pr}
\frac{d^2 P}{dp_x dp_y} = \frac{1}{2\pi} \left | \int d^2 r \left [A_1 (y , \mathbf r -\frac{\mathbf b}{2}) - A_2 (y , \mathbf r +\frac{\mathbf b}{2}) \right ] e^{i\mathbf p \cdot \mathbf r}  \right |^2.
\end{equation}

\begin{figure*}[htbp]
    \centering 

    \begin{minipage}[t]{0.48\linewidth}
        \centering
        \includegraphics[width=0.95\linewidth]{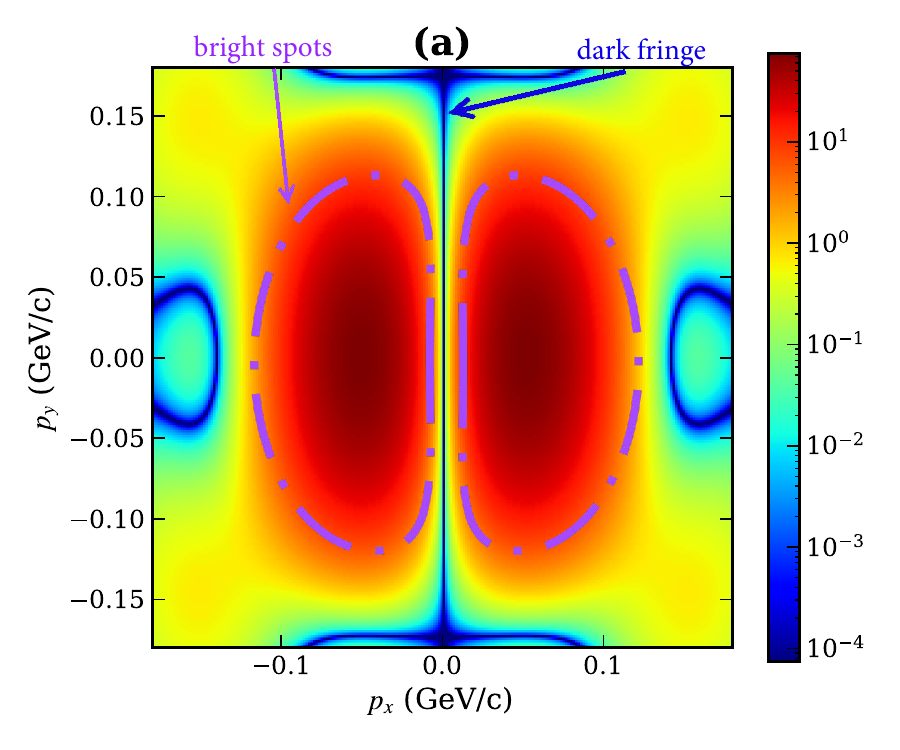}
    \end{minipage}
    \hfill 
    \begin{minipage}[t]{0.48\linewidth}
        \centering
        \includegraphics[width=0.95\linewidth]{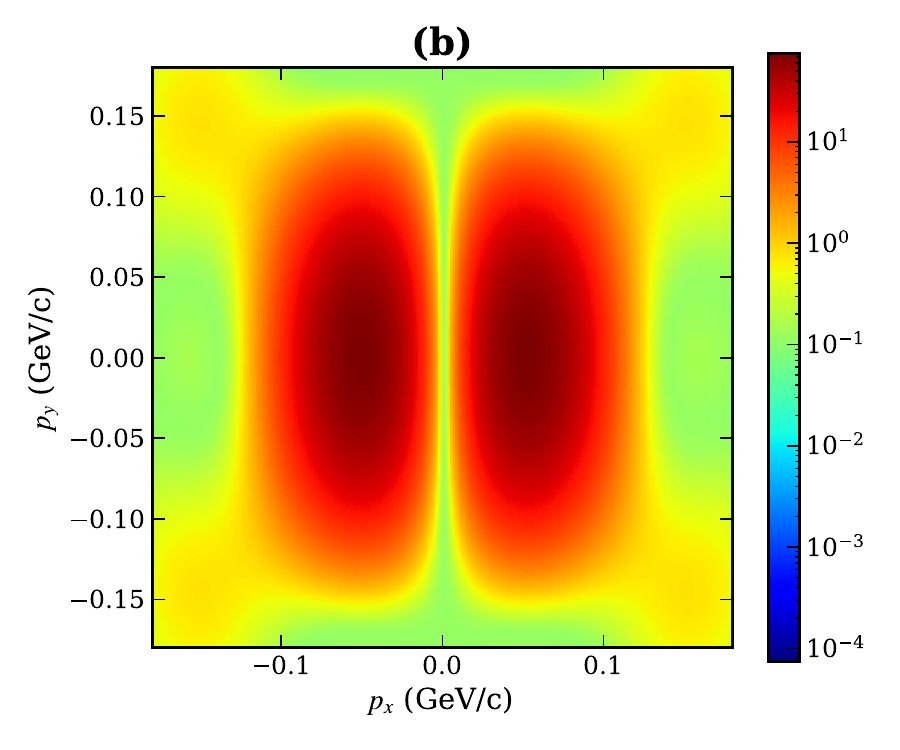}
    \end{minipage}
    \caption{\footnotesize (Color online) 
    Transverse momentum distribution patterns of $J/\psi$ photoproduction in Zr+Zr collisions at mid-rapidity ($y=0$) 
    with an impact parameter of $b = 12$~fm. 
    Panel (a) shows the coherent contribution, characterized by dark interference fringes and bright diffraction spots, while panel (b) includes both coherent and incoherent processes.}
    \label{fig:12_panels}
\end{figure*}

Figure~\ref{fig:12_panels} displays the simulated $J/\psi$ photoproduction probability in the transverse momentum plane at mid-rapidity ($y=0$) with $b=12$~fm. 
Focusing on the coherent process in Fig.~\ref{fig:12_panels}(a), within the transverse momentum range $p_x, p_y \in [-0.17, 0.17]$~GeV/$c$, the distribution exhibits a clear interference pattern characterized by a central dark fringe situated between a pair of bright spots.

Experimental measurements are inherently contaminated by incoherent photoproduction, in which the photon interacts with an individual nucleon and breaks the coherence of the nucleus, often leading to nuclear dissociation. Unlike the coherent process, which acts as a slit-like interferometer, the incoherent process produces a diffuse momentum distribution that fills in the diffractive minima, as shown in Fig.~\ref{fig:12_panels}(b). We model this contribution using the standard Glauber approach~\cite{wang2022calculations,klein2017starlight,klein1999exclusive}, incorporating nuclear thickness functions and photon-nucleon cross sections. The detailed formalism is given in Appendix~\ref{detail}. Our training dataset explicitly includes this incoherent background, forcing the network to learn robust geometric features that survive this degradation. The pattern shown in Fig.~\ref{fig:12_panels} corresponds to just one representative configuration for the tip-tip case with a quadrupole deformation of $\beta_2 = 0.3$. The full set of results, covering all deformation and neutron-skin configurations, is presented in Fig.~\ref{fig:12_panels_ap} of Appendix~\ref{detail}.

While theoretical calculations typically assume fixed nuclear orientations, experimental measurements represent an ensemble average over randomly oriented nuclei. To emulate these conditions, our simulation averages over the full rotational phase space.
 For each sample, we integrate the momentum distributions over a uniform sampling of Euler angles 
$(\alpha, \beta, \gamma)$, thereby capturing the characteristic smearing of diffractive patterns caused by the random spatial alignment of the deformed nuclei.

Furthermore, to match the kinematic acceptance of collider experiments 
and maximize the sensitivity to nuclear structure, we apply specific kinematic cuts. Since coherent $J/\psi$ photoproduction is predominantly concentrated in the very low transverse momentum region, we restrict the input images to the window of 
$p_x, p_y \in [-0.17, 0.17]$~GeV/$c$. This selection isolates the coherent signal from the high-$p_T$ incoherent background.

To represent the nuclear ground state under experimental conditions, every generated image incorporates simultaneous variations of both nuclear deformation ($\beta_2$, $\beta_3$) and neutron skin. Rather than training on isolated effects, our simulation captures the interplay between deformation and neutron-skin effects. This ensures that the generated patterns reflect the entangled signatures expected in real events.

\subsection{Multitask deep-learning framework and performance}

\begin{figure*}[htb]
\centering
\includegraphics[width=0.95\textwidth]{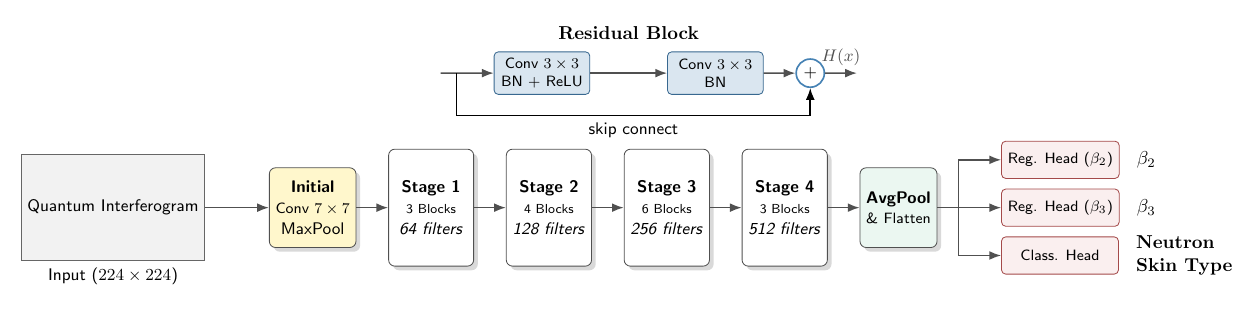}
\caption{Schematic of the Multitask deep-learning architecture.
The pipeline accepts the two-dimensional transverse momentum distribution (Left) and processes it through a ResNet-34 backbone to extract geometric features. 
The Top row details the internal structure of a single Residual Block with a skip connection. 
The Bottom row shows the hierarchical feature extraction stages with increasing filter depths, leading to three task-specific heads (Right) for parameter regression and classification.
\footnotesize \textbf{Conv}: Convolutional layer for local feature extraction.
\textbf{BN}: Batch Normalization. \textbf{ReLU}: Non-linear activation function.
\textbf{MaxPool}: Max pooling. \textbf{Filters}: Number of feature channels in each stage.
\textbf{$H(x)$}: Block output summing the learned mapping and identity.
\textbf{AvgPool}: Global average pooling. \textbf{Flatten}: Reshaping features from
two-dimensional maps to a one-dimensional vector. \textbf{Reg/Class Head}: MLP for
final physical parameter regression or classification.}
\label{fig:resnet_arch}
\end{figure*}

The architecture of our framework is illustrated in Fig.~\ref{fig:resnet_arch}. We interpret the two-dimensional transverse momentum distribution as a quantum interferogram, a phase-sensitive pattern in which the nuclear geometry is encoded in the positions of the diffraction minima and interference fringes. To decode these high-frequency structures, we employ a ResNet-34 backbone~\cite{he2016deep}. The residual identity mappings preserve the sharp intensity gradients associated with diffraction nodes, limiting the loss of this fine-scale information as features propagate through deep layers. This keeps the network sensitive to the interference details where the deformation signatures are most strongly imprinted, rather than only the overall envelope.

To resolve the final-state degeneracy between neutron skin and deformation, our framework adopts a Multitask learning strategy. In conventional analyses, global fits to integrated observables such as $v_n(p_T)$ average over kinematic regions with distinct physical sensitivities, causing these two structural effects to become difficult to distinguish. By contrast, our design shares the feature-extraction backbone (Stages 1--4) before branching into task-specific heads. This forces the network to learn a unified latent representation that captures the joint dependence of the interferogram on both effects. By modeling the correlated response in the shared layers and decoupling it only at the output stage, the framework separates the diffraction-dominated skin signal from the interference-dominated shape signal, resolving the parameter ambiguities of standard fitting procedures.

\begin{figure*}[htb]
\hspace*{-21mm}
\begin{minipage}[t]{69mm}
\includegraphics[width=0.9\textwidth]{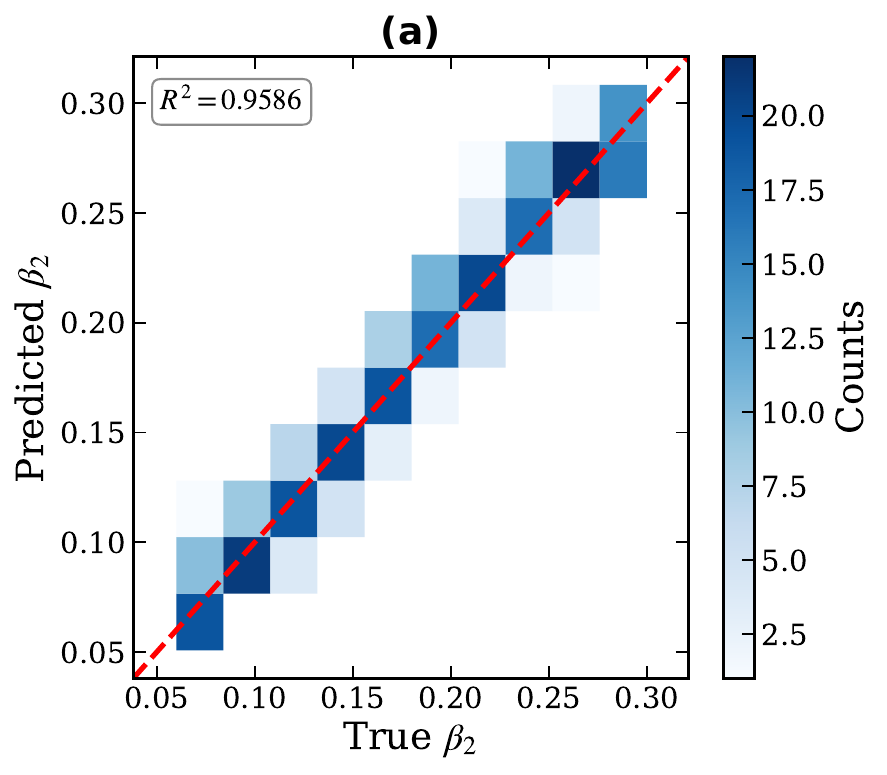}
\end{minipage}
\hspace{-7mm}
\begin{minipage}[t]{69mm}
\includegraphics[width=0.9\textwidth]{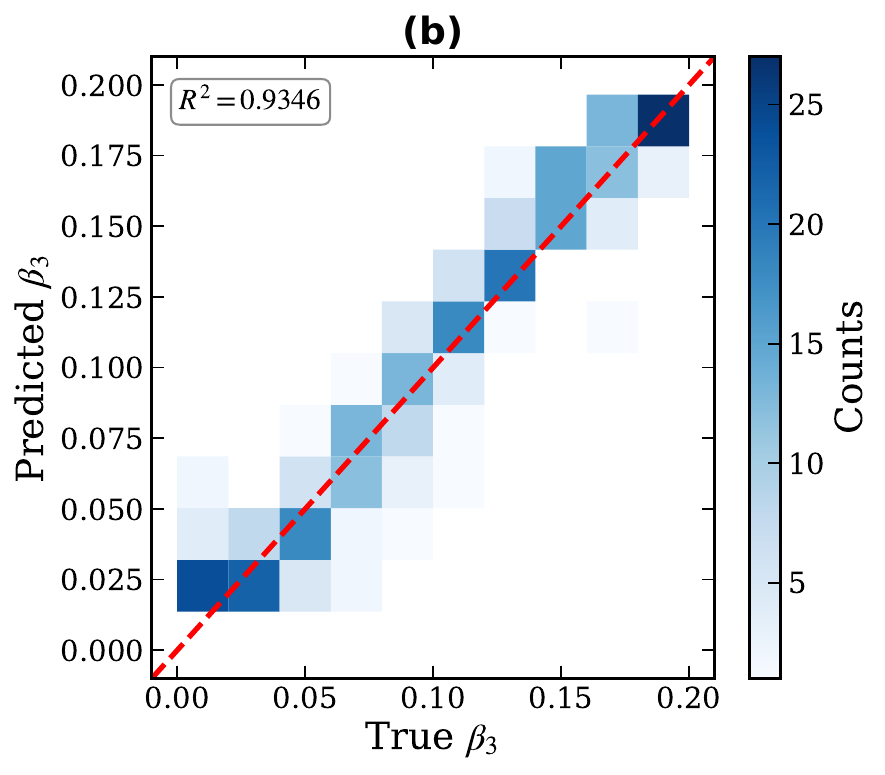}
\end{minipage}
\hspace{-7mm}
\begin{minipage}[t]{69mm}
\includegraphics[width=0.9\textwidth]{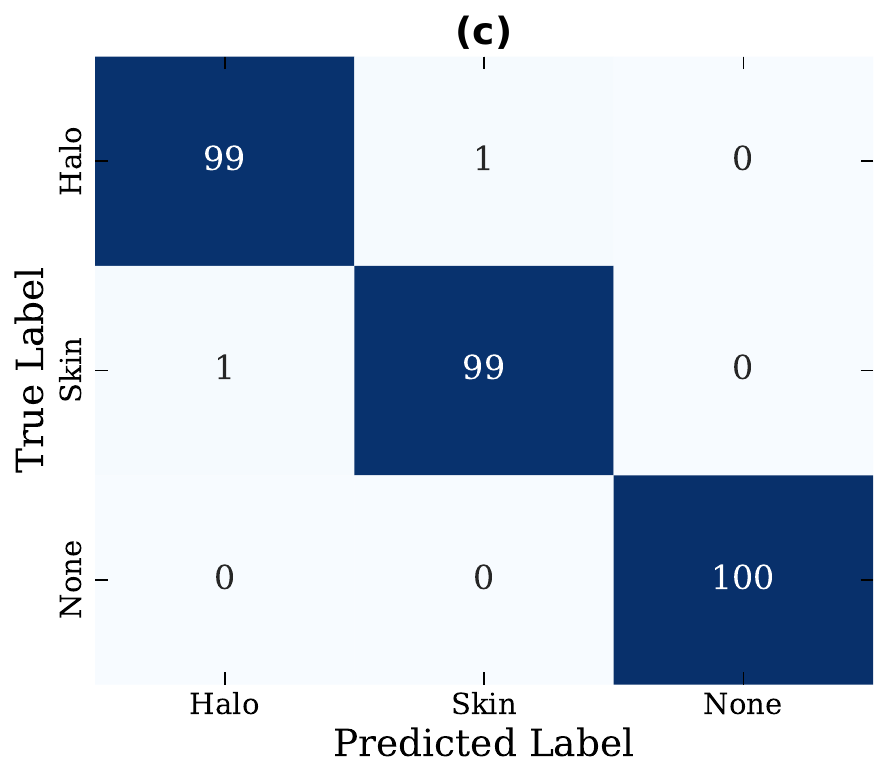}
\end{minipage}
\hspace*{-17mm}
\caption{\footnotesize(Color online) Two-dimensional histograms comparing model predictions with ground-truth values using coherent + incoherent input. The plots show the regression results for $\beta_2$ (a) and $\beta_3$ (b), and the confusion matrix for neutron-skin type classification (c).} \label{fig:coh_result}
\end{figure*}

\begin{table}[htbp]
  \centering
  \setlength{\tabcolsep}{12pt} 
  \caption{Performance comparison under different noise levels ($n$). Here, $n=1$ corresponds to the realistic experimental incoherent background level derived from the Glauber model [as shown in Fig.~\ref{fig:12_panels}(b)].}\label{tab:noice_performance}
  \begin{tabular}{cccc} 
    \toprule
     \hline
    Noise ($n$) & $R^2(\beta_2)$ & $R^2(\beta_3)$ & Acc (Skin) \\
     \hline
    \midrule
    0           & 0.9727         & 0.9445         & 0.9833     \\
    1           & 0.9586         & 0.9346         & 0.9933     \\
    100     	& 0.7787         & 0.7235 	      & 0.8500
   \\
    \bottomrule
     \hline
  \end{tabular}
\end{table}

Moreover, strictly coherent photoproduction represents an idealized limit. In realistic measurements, incoherent photo-nucleus interactions generate diffuse backgrounds that suppress and broaden diffraction minima, as shown in Fig.~\ref{fig:12_panels}. This contamination further entangles the structural parameters by eroding phase-sensitive information.

To test whether the architecture can operate under these conditions, we evaluate the model on datasets that explicitly include incoherent contributions. Figure~\ref{fig:coh_result} shows the regression and classification accuracy obtained with the incoherent background included. 

To rigorously quantify the survival of these geometric signatures, we performed a stress test by scaling the incoherent cross section by a factor $n$, where $n=1$ corresponds to the standard theoretical expectation derived from our model. The robustness of the framework is summarized in Table~\ref{tab:noice_performance}. The predictive accuracy for both deformation parameters ($\beta_2$, $\beta_3$) and the neutron-skin type remains stable even as the background is amplified by nearly two orders of magnitude ($n\sim 10^2$). At $n>10^2$, however, the accuracy for $\beta_3$ degrades more rapidly than that for $\beta_2$. As detailed in Appendix~\ref{sec:deformation_ratio}, this behavior originates from the different manifestations of the deformations in momentum space: $\beta_2$ induces a global, high-contrast distortion of the dark-fringe pattern, whereas $\beta_3$ leads to more subtle, localized modifications to the interference fringes. Consequently, the fine-grained features associated with $\beta_3$ are more susceptible to being masked by the diffuse incoherent background than the gross structural changes driven by $\beta_2$.

The performance degrades significantly only when the background completely overwhelms the coherent signal. This margin ensures that the method is robust against theoretical uncertainties in modeling the incoherent background and against experimental imperfections in background subtraction.


\section{INTERPRETABILITY AND PHYSICAL VALIDATION}\label{sec:interpre}

\begin{figure*}[htbp]
    \centering 

    \begin{minipage}[t]{0.48\linewidth}
        \centering
        \includegraphics[width=0.95\linewidth]{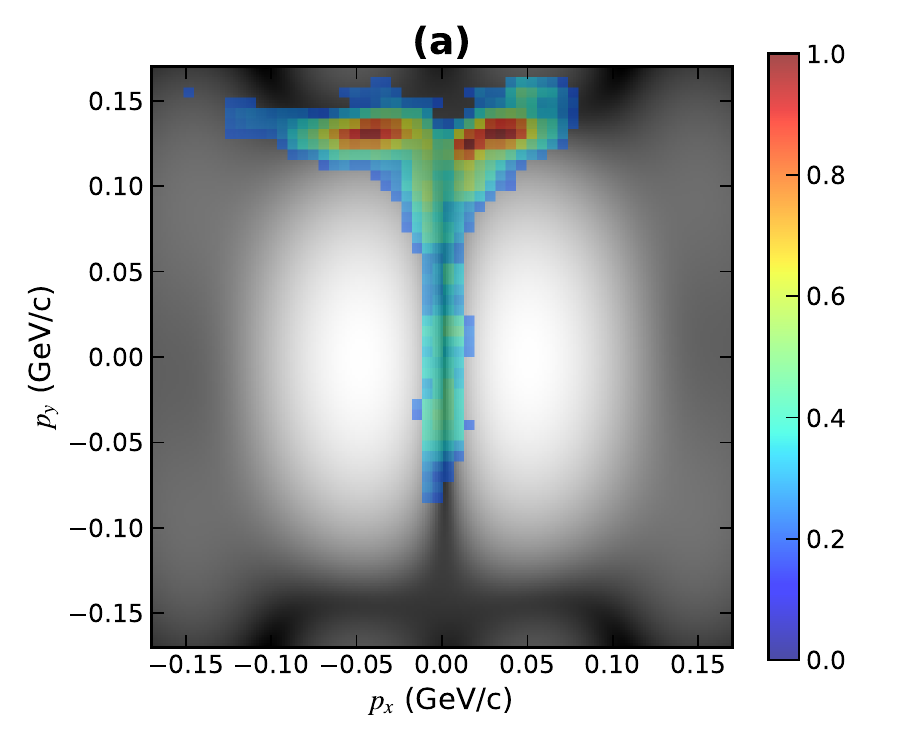}
    \end{minipage}
    \hfill 
    \begin{minipage}[t]{0.48\linewidth}
        \centering
        \includegraphics[width=0.95\linewidth]{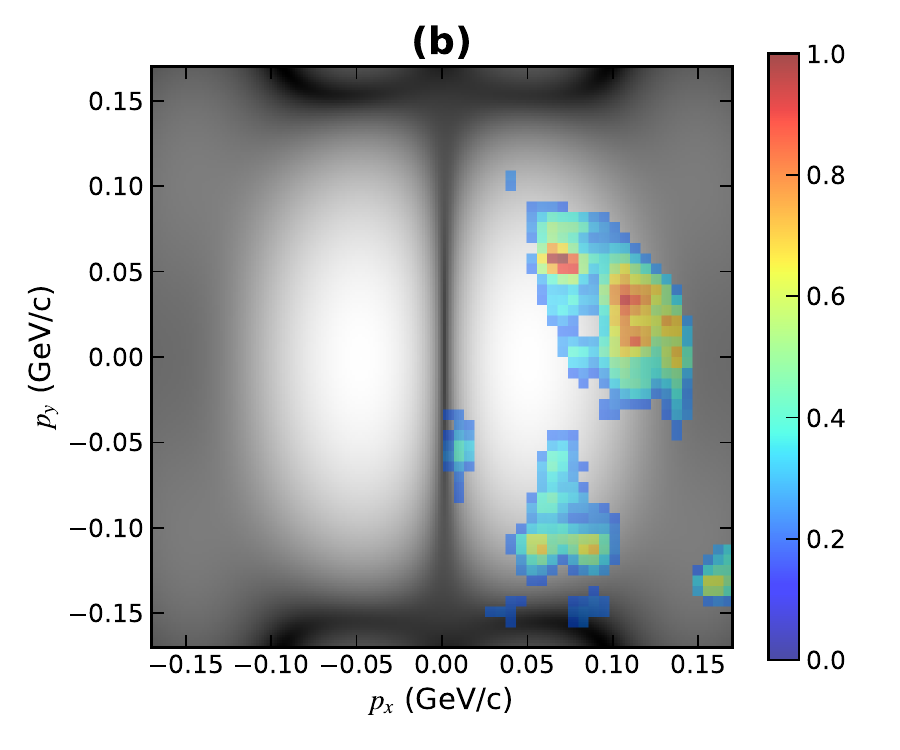}
    \end{minipage}
    
    \caption{\footnotesize (Color online) Interpretability analysis using PDA. 
    (a) For nuclear deformation prediction ($\beta_2, \beta_3$), the model focuses on the dark fringes.
    (b) For neutron-skin classification, the model attention shifts to bright spots. }
    \label{fig:pda}
\end{figure*}

To interpret the model's predictions, 
we apply Prediction Difference Analysis (PDA)~\cite{zintgraf2017visualizing}. 
This technique evaluates the sensitivity of the network's output to localized variations in the input image, thereby identifying the specific regions of phase space that carry the most critical information for each physical parameter.

Without any prior knowledge of the underlying formalism, the network learns to isolate distinct physical drivers of the coherent photoproduction cross section. As shown in Fig.~\ref{fig:pda}, different prediction tasks are associated with clearly separated regions in transverse momentum space. 
The regression of deformation parameters $\beta_2$ and $\beta_3$ is driven by the dark fringes [Fig.~\ref{fig:pda}(a)], corresponding to the interference fringes of the pattern, reflecting the phase-shift effect of Eq.~(\ref{Eq:def}).
In contrast, neutron-skin classification relies primarily on the edges of the bright spots [Fig.~\ref{fig:pda}(b)], corresponding to the diffraction-amplitude modulations described by Eq.~(\ref{Eq:skin}).
These complementary sensitivities indicate that the network distinguishes between phase-sensitive and amplitude-sensitive structures in the interference pattern.

\begin{figure*}[htb]
\hspace*{-21mm}
\begin{minipage}[t]{69mm}
\includegraphics[width=0.9\textwidth]{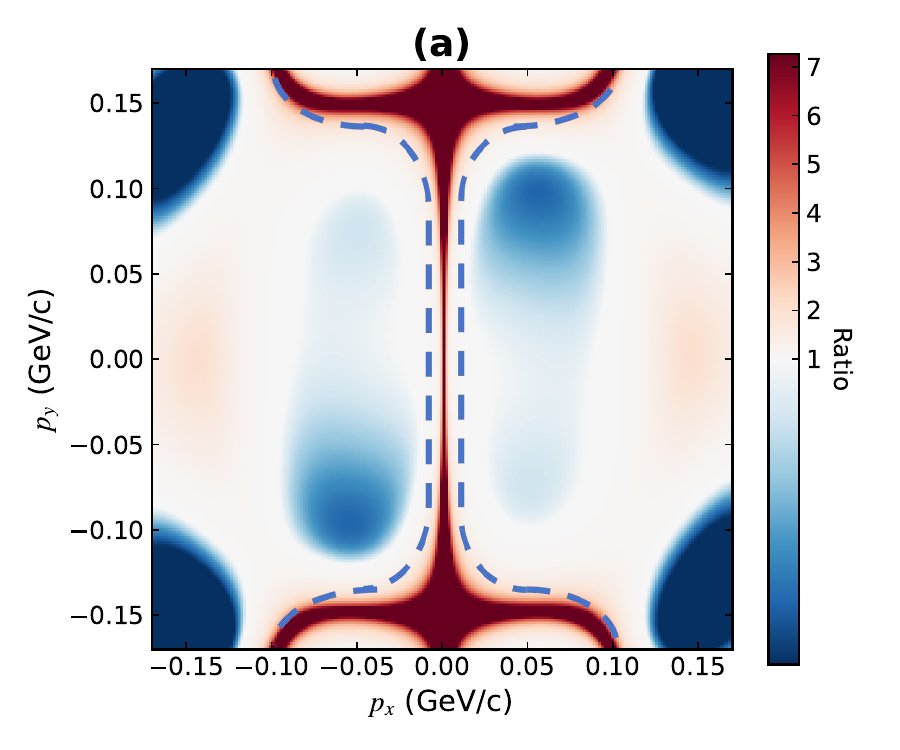}
\end{minipage}
\hspace{-7mm}
\begin{minipage}[t]{69mm}
\includegraphics[width=0.9\textwidth]{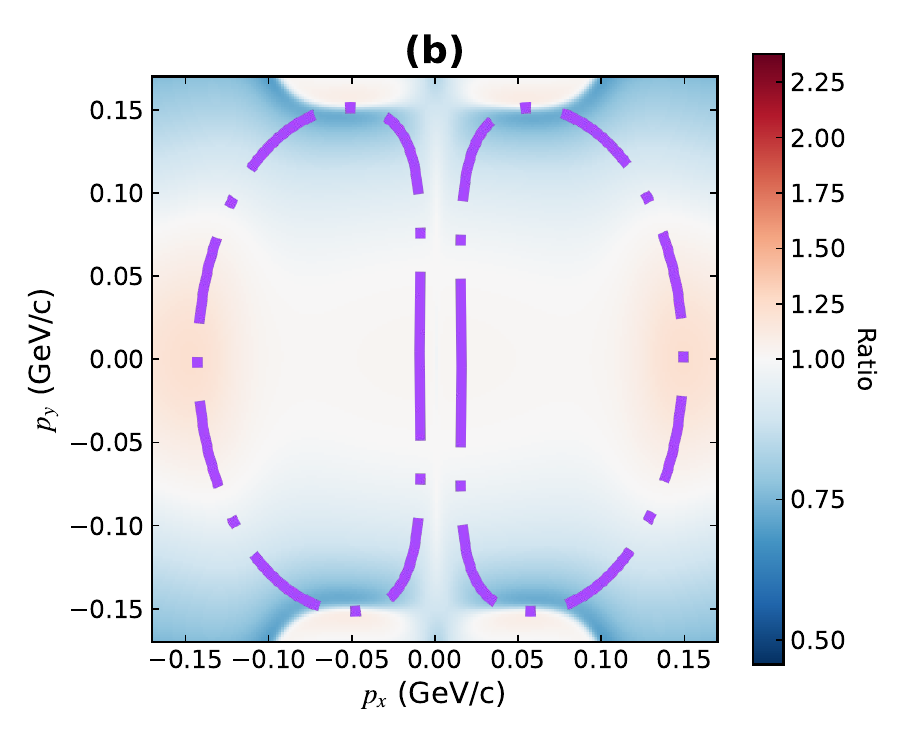}
\end{minipage}
\hspace{-7mm}
\begin{minipage}[t]{69mm}
\includegraphics[width=0.9\textwidth]{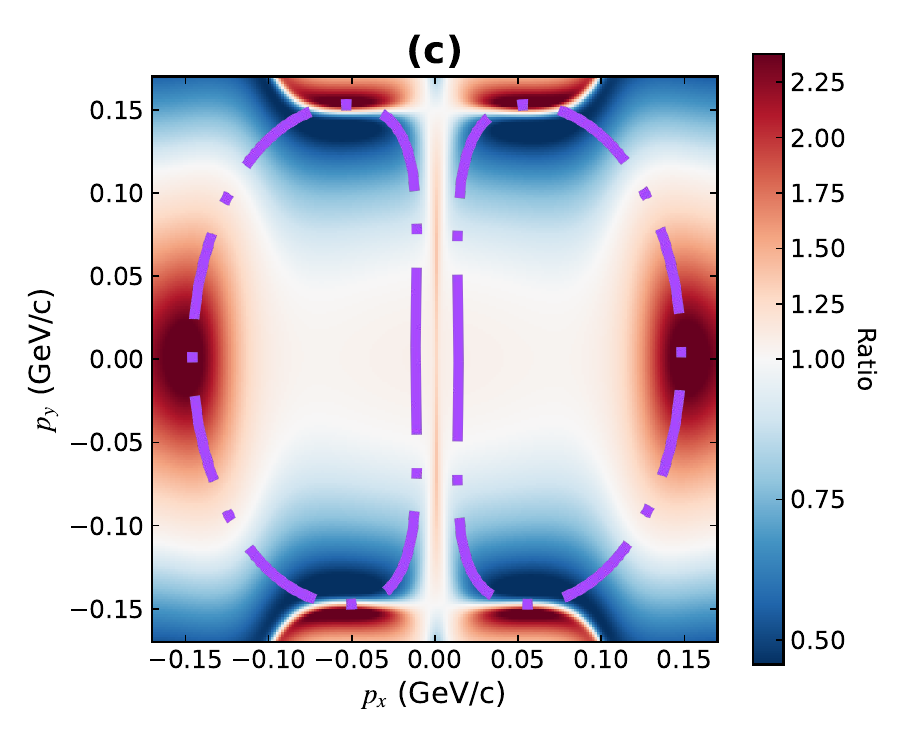}
\end{minipage}
\hspace*{-17mm}
\caption{\footnotesize(Color online) 
Ratios of average transverse momentum distributions. Panel (a): the ratio of the highly deformed case (maximum $\beta_2$ and $\beta_3$ in the dataset) relative to the spherical nucleus; the blue dashed lines mark the central dark-fringe region. Panels (b) and (c): the ratios of halo-type and skin-type neutron-skin cases relative to the no-skin baseline; the violet dashed contours outline the pair of bright spots.
} \label{fig:ratio}
\end{figure*}

This conclusion is supported by the momentum-distribution ratios shown in Fig.~\ref{fig:ratio}.
Comparing highly deformed nuclei with spherical ones [Fig.~\ref{fig:ratio}(a)], the relative yield difference is largest at the interference fringes, showing that deformation information is encoded in the filling of the dark fringes. Figures~\ref{fig:ratio}(b) and \ref{fig:ratio}(c) show the momentum-distribution ratios for halo- and skin-type nuclei relative to a no-skin baseline; the differences are concentrated around the edges of the interference bright spots.

\section{AI-INSPIRED OBSERVABLES AND EXPERIMENTAL SIGNATURES}\label{sec:obser}

\subsection{Image-based disentanglement observables}

\begin{figure*}[htb]
\centering
\begin{minipage}[c]{0.9\textwidth} 
    \centering
    \includegraphics[width=0.9\textwidth]{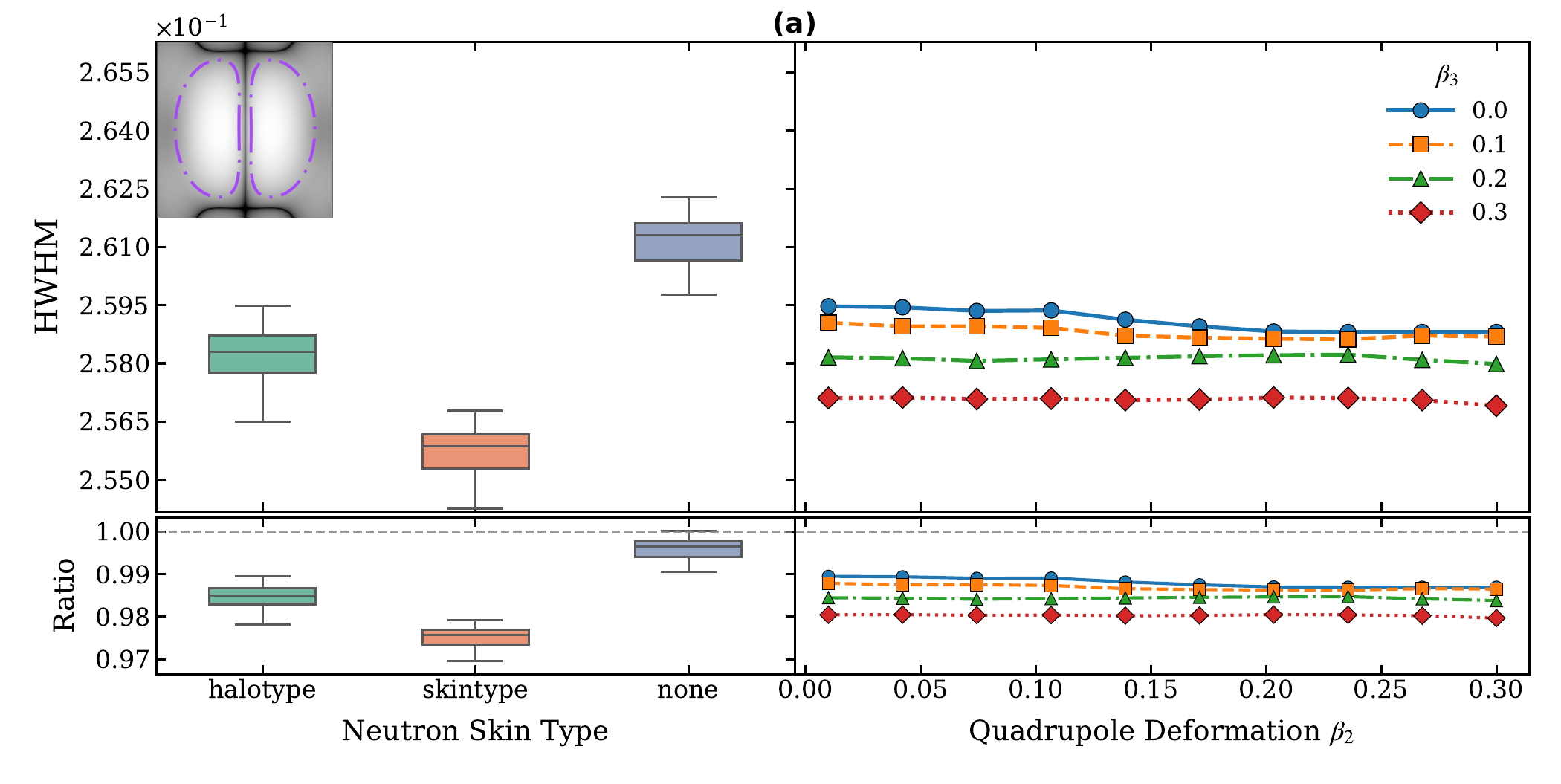} \\
    \includegraphics[width=0.9\textwidth]{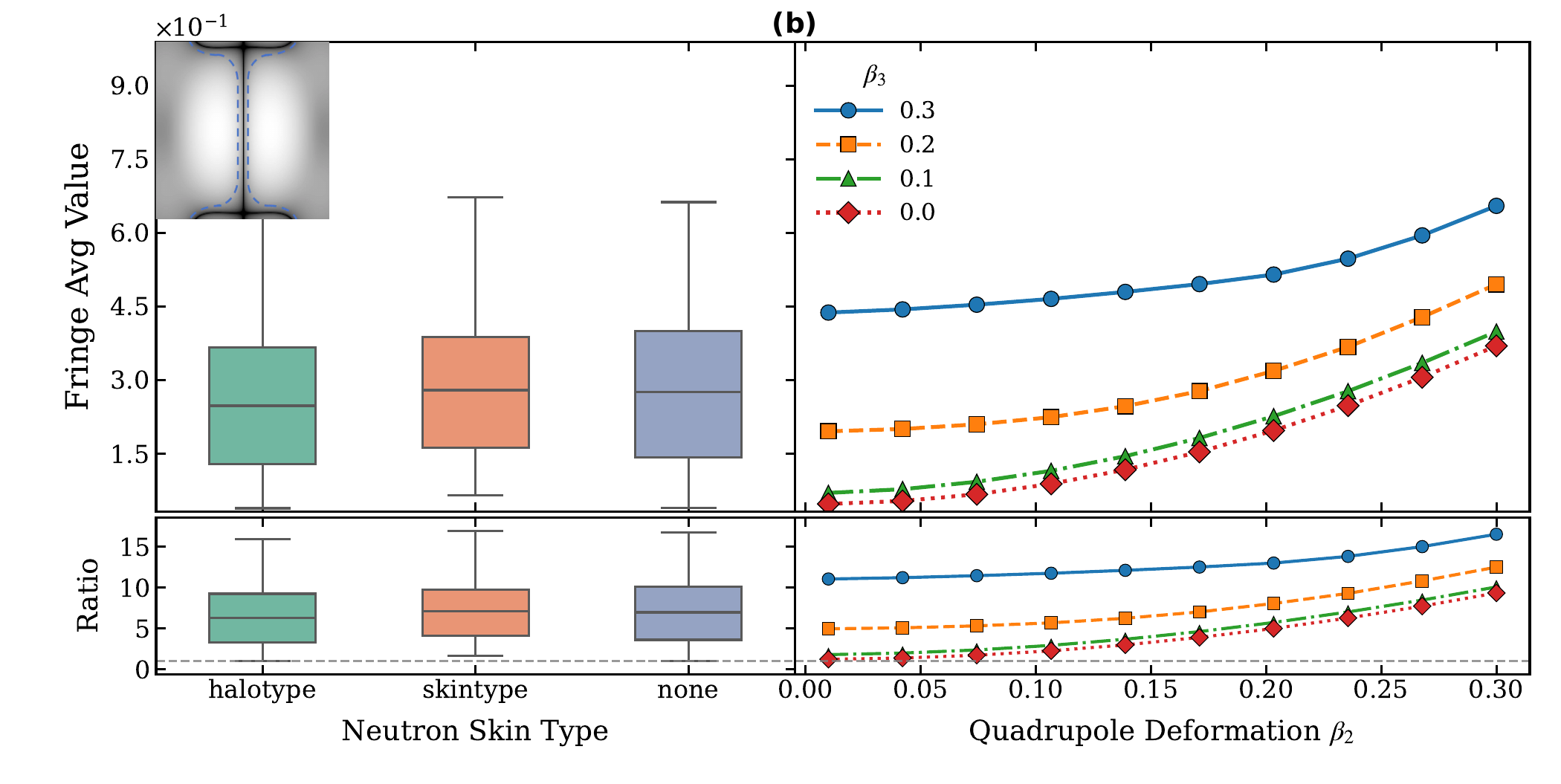}
\end{minipage}
\caption{\footnotesize (Color online) 
Definition and sensitivity of the observables. 
 The insets show the momentum distribution regions corresponding to the definitions of the HWHM (a) and fringe average value (b). The dashed violet contours mark the high-intensity regions (bright spots) defining the HWHM, while the dashed blue contours indicate the diffraction minimum used for the fringe average value.
(a) dependence of the HWHM on neutron-skin type and deformation ($\beta_2$, $\beta_3$).
(b) fringe average value on the same parameters. 
} 
\label{fig:AD}
\end{figure*}

The interpretability analysis in Sec.~\ref{sec:interpre} reveals that the deep-learning model disentangles nuclear structure by attending to distinct spatial regions: the bright diffraction peaks for the neutron skin and the dark interference fringes for the deformation. Guided by these attention maps, we introduce two image-based observables that mimic the network's focus. These quantities encode the spatial patterns identified by the network and connect them to kinematic observables measurable in current heavy-ion experiments.

The first, the half-width at half-maximum (HWHM), is defined as the integrated area where $I > I_{\text{max}}/2$. This metric characterizes the area of the bright spots and is primarily sensitive to the neutron skin, remaining robust against $\beta_2$ and $\beta_3$ variations [Fig.~\ref{fig:AD}(a)]. 
This decoupled sensitivity lets the HWHM serve as a clean proxy for the radial mass distribution, isolating it from angular asymmetries.

Nuclear deformation introduces orientation-dependent phase shifts that fill in the interference minima. The neural network identifies this effect in the dark-fringe regions. We therefore define the fringe average value as the mean intensity of the lowest 5\% of pixels. As shown in Fig.~\ref{fig:AD}(b), this value depends monotonically on the deformation parameters $\beta_2$ and $\beta_3$ while remaining nearly independent of the neutron skin. Physically, it measures the interference visibility: a spherical nucleus produces deep, sharp minima (low fringe average value ), whereas strong deformation smears these minima through orientation averaging (high fringe average value ).

\subsection{Connection to experimental observables}

While the image-based observables offer the cleanest separation of nuclear structure, they require high-resolution two-dimensional imaging, which is experimentally challenging. The physical insight gained from these morphological features instead points to suitable kinematic windows for more accessible observables. In this section, we show that integrated observables in specific kinematic windows capture the essential structural information, providing a practical strategy for experiments with limited statistics. The HWHM characterizes the area of the bright spots, so a larger HWHM corresponds to a momentum distribution shifted toward higher values, i.e., an increase in $\langle p_T \rangle$ within the range $p_T < 0.17$~GeV/$c$.

\begin{figure*}[htb]
\hspace*{-21mm}
\begin{minipage}[t]{69mm}
\includegraphics[width=0.9\textwidth]{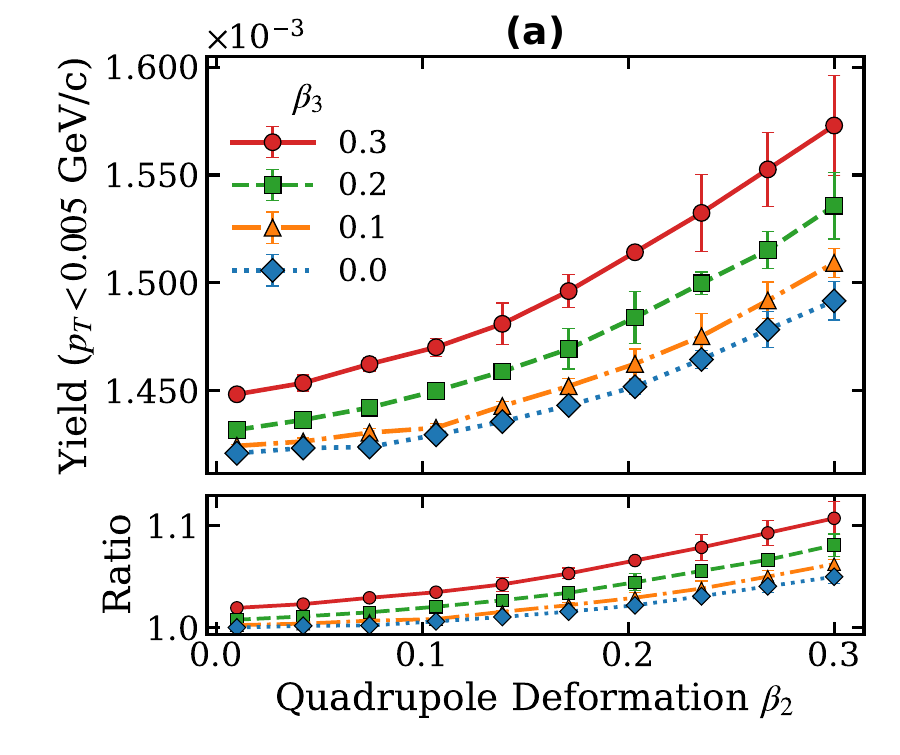}
\end{minipage}
\hspace{-7mm}
\begin{minipage}[t]{69mm}
\includegraphics[width=0.9\textwidth]{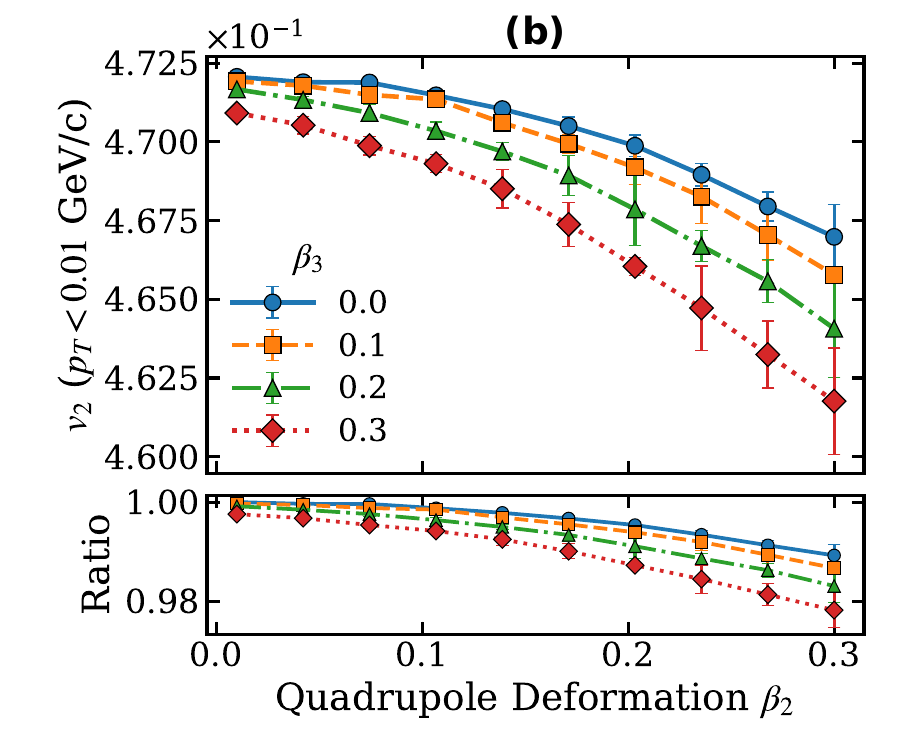}
\end{minipage}
\hspace{-7mm}
\begin{minipage}[t]{69mm}
\includegraphics[width=0.9\textwidth]{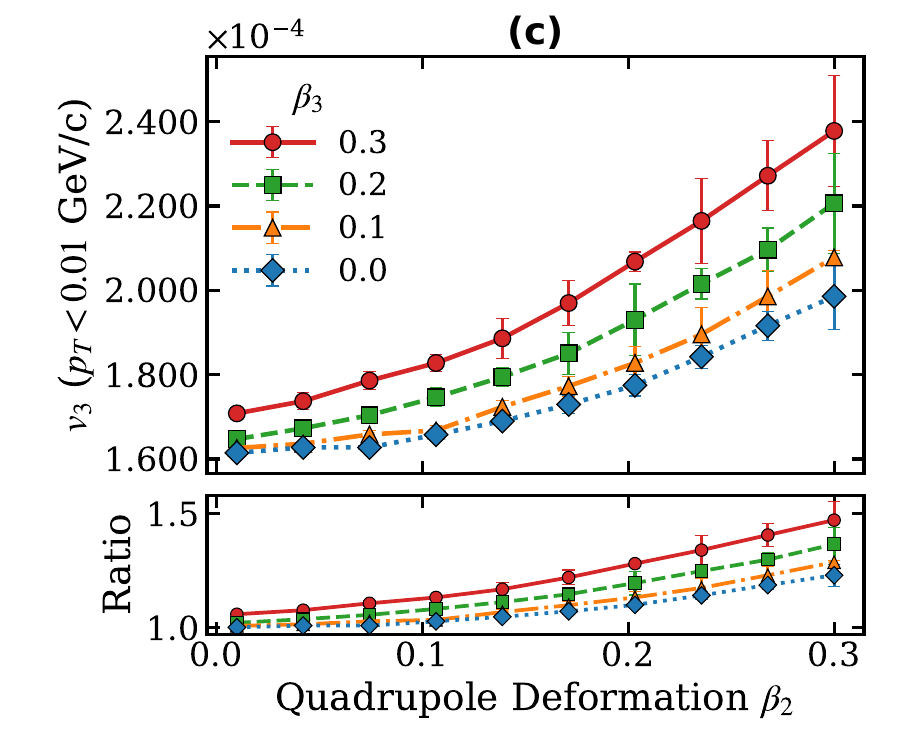}
\end{minipage}
\hspace*{-17mm}
\caption{\footnotesize(Color online) Physical observables sensitive to nuclear deformation, corresponding to the image-based fringe average feature. The plots show the yield ratio (a), 
elliptic flow $v_2$ (b), and triangular flow $v_3$ (c) in the low-$p_T$ region ($p_T < 0.005$~GeV/$c$) 
as functions of quadrupole deformation $\beta_2$. 
The lower panels show the ratio of each observable relative to the spherical nucleus case.} \label{fig:ob_def}
\end{figure*}

The fringe average value quantifies the filling of the interference minima. 
This effect is most prominent in the low-$p_T$ regime ($p_T < 0.005$ ~GeV/$c$), 
where the primary interference fringes reside. In this region, deformation-induced phase smearing produces 
monotonic variations in $v_2$ and $v_3$ as functions of $\beta_2$ and $\beta_3$ (Fig.~\ref{fig:ob_def}),
as well as in the yield ratio, defined as the ratio of the yield for $p_T < 0.005$~GeV/$c$ to that for $p_T < 0.17$~GeV/$c$.
The low-$p_T$ yield ratio and anisotropic flow therefore provide sensitive probes of nuclear deformation and serve as the experimental counterparts of the fringe average value.

\begin{figure*}[htb]
\hspace*{-21mm}
\begin{minipage}[t]{69mm}
\includegraphics[width=0.9\textwidth]{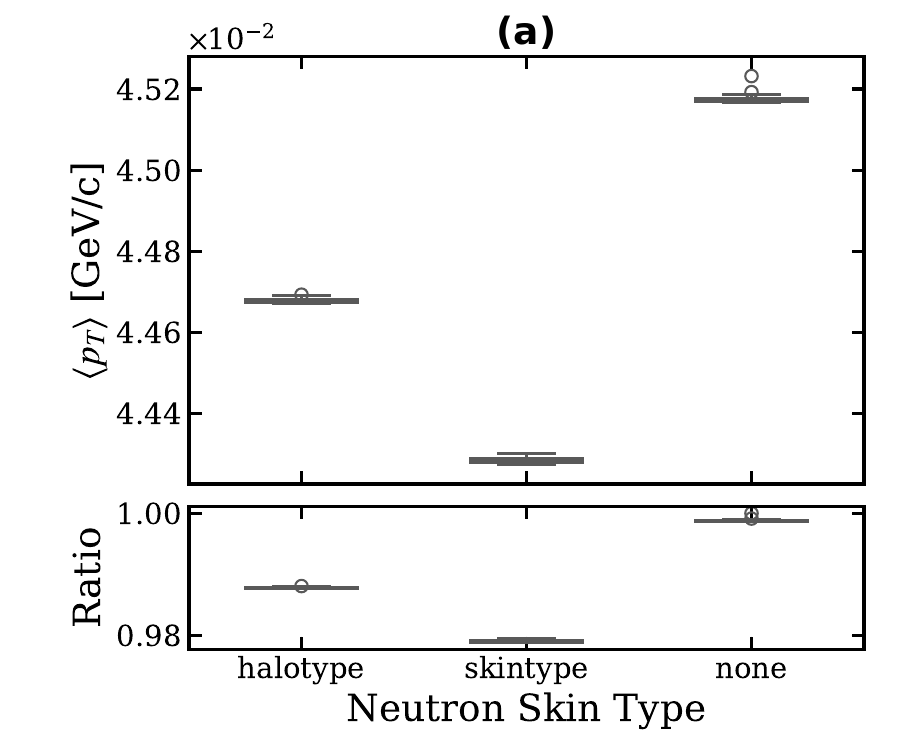}
\end{minipage}
\hspace{-7mm}
\begin{minipage}[t]{69mm}
\includegraphics[width=0.9\textwidth]{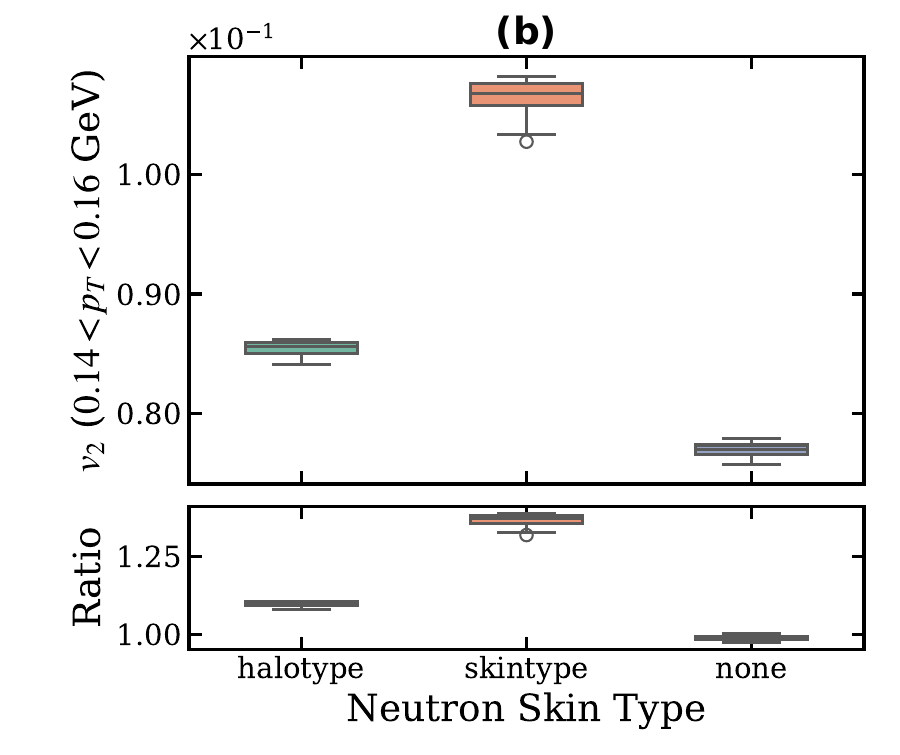}
\end{minipage}
\hspace{-7mm}
\begin{minipage}[t]{69mm}
\includegraphics[width=0.9\textwidth]{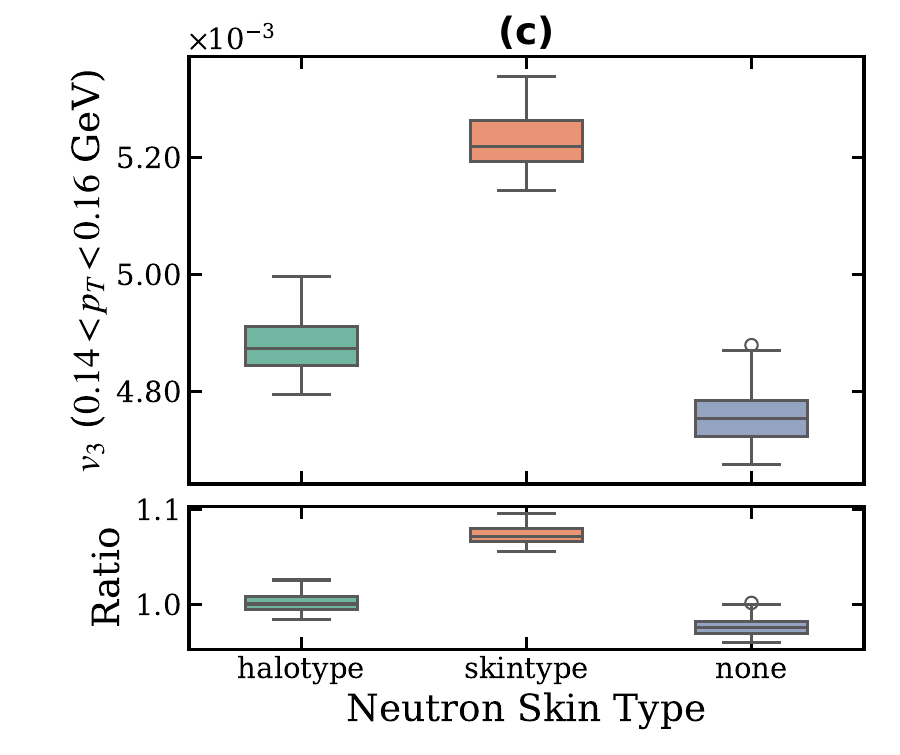}
\end{minipage}
\hspace*{-17mm}
\caption{\footnotesize(Color online) Physical observables sensitive to neutron-skin type, 
which are the physical counterparts of the image-based HWHM feature. 
The panels display the mean transverse momentum $\langle p_T \rangle$ calculated within the range $p_T < 0.17$~GeV/$c$ (a), 
and the anisotropic flow coefficients $v_2$ (b) and $v_3$ (c) 
calculated in the relatively high-$p_T$ region ($0.14 < p_T < 0.16$~GeV/$c$). 
} \label{fig:ob_skin}
\end{figure*}

In contrast, the high-$p_T$ region ($0.14 < p_T < 0.16$~GeV/$c$) aligns with the edges of the bright spots, 
where neutron-skin effects predominate. 
Simulations confirm that different neutron-skin structures significantly modulate $v_2$ and $v_3$ in this kinematic range (Fig.~\ref{fig:ob_skin}). 
Thus, anisotropic flow at relatively high $p_T$ provides a distinct signature for characterizing the nuclear neutron skin.

In summary, our analysis reduces the high-dimensional correlations learned by the network to a small set of interpretable signatures. Although the deep-learning model uses the full two-dimensional momentum distribution, we show that integrated observables in specific kinematic windows capture the essential structural information. This establishes a data-driven protocol for future experiments: constraints on nuclear geometry can be extracted by targeting these kinematic windows, even with limited statistics.

\section{CONCLUSION}\label{sec:conclu}

In this work, we have developed a Multitask deep-learning framework for quantum-interference tomography of the atomic nucleus, treating ultra-peripheral collisions as femtoscopic double-slit experiments. Applying this framework to the diffractive patterns of $^{96}_{40}\text{Zr}$, we show that the entangled signatures of nuclear shape (deformation) and surface diffuseness (neutron skin) can be effectively decoupled. The ability of the model to resolve this holographic degeneracy, in which different geometric configurations produce nearly identical conventional observables, illustrates the value of interpreting high-energy nuclear data through the wave-optics AI-enhanced lens. This approach turns the transverse momentum distribution into a quantitative probe of nuclear structure,  isolating subtle nuclear structure effects from noisy interference patterns even in heavy nuclei.

A central practical contribution of this study is the translation of deep-learning features into experimentally accessible observables. Direct application of image-based networks to experimental data is often limited, because measuring high-resolution two-dimensional distributions requires very high statistical precision. Our framework addresses this by mapping the sensitivity patterns identified by the network onto conventional physical quantities within specific kinematic windows. We demonstrate that the neutron-skin sensitivity corresponds to the mean transverse momentum and anisotropic flow coefficients calculated in the high transverse momentum region. Conversely, deformation effects are mapped to the anisotropic flow coefficients and the yield ratio in the low transverse momentum interference minima. 
These findings provide a practical guide for future experiments at the LHC~\cite{grundRecentResultsUltraperipheral2025,massacrier2024coherent}, RHIC~\cite{aboona2024observation}, and the future Electron-Ion Collider (EIC)~\cite{Accardi:2012qut}. In particular, our results suggest that competing geometric effects could be disentangled using standard flow harmonics and yield ratios within the proposed kinematic windows, without requiring full image reconstruction or model-dependent global fits.

This deep-learning-assisted interferometry suggests several further directions. Beyond the ground-state geometry, it offers a route to image the spatial gluon density~\cite{BRANDENBURG2025104174}, probing nuclear parton distribution functions~\cite{guzey2013evidence} and generalized parton distributions~\cite{garcon2003introduction}
with high spatial resolution~\cite{mantysaari2016evidence}. The framework extends naturally to other vector mesons such as $\rho^0$~\cite{shen2024exploring} and to higher-dimensional observables such as the polarization-dependent decay angles of $J/\psi$, which could provide independent constraints on initial-state fluctuations~\cite{shao2025geometryinducedazimuthalanisotropycoherent,alice2023first,zhao2024j}.
Combined with Bayesian inference~\cite{pavone2023machine,bernhard2016applying}, the approach could also yield posterior constraints on the nuclear equation of state~\cite{he2021machine,Pratt:2015zsa}. More broadly, this work connects machine-learning features to the physical principles of quantum interference~\cite{Carleo:2019ptp} and uses them to guide the design of experimental observables. In the coming high-luminosity era, such data-driven tomographic techniques can help decode the fluctuating geometry of the nucleus from the diffractive interference patterns it produces.

\section*{ACKNOWLEDGMENTS}
We thank the DEEP-IN working group at RIKEN-iTHEMS for support in the preparation of this paper.
This work is partially supported by the National Natural Science Foundation of China under Grants No. 12325507, No. 12547102, and No. 12147101, and the National Key Research and Development Program of China under Grant No. 2022YFA1604900 (J.Z. and G.M.). L.W. is supported by the JSPS KAKENHI Grant No. 25H01560, and JST-BOOST Grant No. JPMJBY24H9. W.Z. is supported by Anhui Provincial Natural
Science Foundation No. 2208085J23 and Youth Innovation Promotion Association of Chinese Academy of Sciences.

\section*{DATA AVAILABILITY}
The code, trained model weights, and datasets that support the findings of this study will be made openly available upon acceptance of the article.


\bibliography{Reference.bib}

\newpage
\appendix
\section{GLAUBER-VMD FRAMEWORK DETAIL}\label{detail}

$J/\psi$ photoproduction in $^{96}_{40}\text{Zr}$+$^{96}_{40}\text{Zr}$ ultra-peripheral collisions is simulated using the Glauber-VMD framework
established in Refs.~\cite{shen2024exploring,luo2023effect,zha2019double}.
The spatial distribution of nucleons inside the $^{96}_{40}\text{Zr}$ nucleus is modeled
with a deformed Woods--Saxon profile:

\begin{equation}
\rho_A(r,\theta,\phi) = \frac{\rho^0}{1+\exp[(r-R(\theta,\phi))/a]},
\end{equation}

where the radius parameter $R(\theta,\phi)$ incorporates deformations:

\begin{equation}
R(\theta,\phi) = R_0\left [ 1+\beta_2 Y_{2,0}(\theta , \phi) +\beta_3 Y_{3,0}(\theta , \phi)   \right ].
\end{equation}

Here, $\rho^0$ is the normal nuclear density, 
$R_0$ is the nuclear radius, 
$a$ is the surface diffuseness parameter, 
and $\beta_2$ and $\beta_3$ represent the quadrupole and octupole deformation parameters, respectively.

To model the neutron-skin effect, we distinguish between the proton and neutron density distributions. Specifically, a skin-type neutron skin is implemented by assigning a larger radius to neutrons than protons ($R_{0}^n>R_0^p$) while keeping the diffuseness $a$ identical. Conversely, a halo-type structure is characterized by a larger surface diffuseness for neutrons ($a_n>a_p$) with similar radii.

The coherent production amplitude is formulated by combining the equivalent photon flux with the nuclear scattering amplitude. 
The spatial photon flux induced by the projectile ion is given by the equivalent photon approximation~\cite{krauss1997photon}:

\begin{equation}
\frac{d^3 N_\gamma (\omega _ \gamma , \mathbf x_{\perp})}{d\omega_\gamma d\mathbf x _{\perp}} = \frac{4Z^2 \alpha}{\omega_{\gamma}} \left |\int \frac{d^2 \mathbf k _{\gamma \perp}}{(2\pi)^2} \mathbf k_{\gamma \perp} \frac{F_{\gamma}(\mathbf k_\gamma)}{|\mathbf k_\gamma|^2} e^{i\mathbf x_{\perp}\cdot \mathbf k_{\gamma \perp}}  \right | ^2.
\end{equation}

In this expression, 
$\mathbf{x}_{\perp}$ and $\mathbf{k}_{\gamma\perp}$ are the transverse photon position and momentum vectors, 
$Z$ is the nuclear charge, and $F_{\gamma}(\mathbf{k}_{\gamma})$ is the nuclear electromagnetic form factor.

The scattering amplitude $\Gamma_{\gamma A \to V A}$ is calculated within the Glauber-VMD framework~\cite{miller2007glauber,bauer1978hadronic}, 
accounting for nuclear shadowing:

\begin{equation}
\Gamma_{\gamma A \to V A}(\mathbf{x}_\perp) = \frac{4\sqrt{\alpha}C}{f_V} \frac{f_{\gamma N \to V N}(0)}{f_{\gamma N \to V N}} \times 2 \left[ 1 - \exp\left( -\frac{\sigma_{V N}}{2} T'(\mathbf{x}_\perp) \right) \right],
\end{equation}

with the vector meson-nucleon cross section defined as:

\begin{equation}
\sigma_{V N} = \frac{f_V}{4\sqrt{\alpha} C} f_{\gamma N \to V N}.
\end{equation}

Here, $f_V$ is the vector meson-photon coupling constant, 
and $C$ corrects for non-diagonal couplings~\cite{hufner1998j}. 
To account for coherence effects along the beam direction, we employ $T'(\mathbf{x}_\perp)$:

\begin{equation}
T'(\mathbf{x}_\perp) = \int_{-\infty}^{+\infty} \rho_A(\sqrt{\mathbf{x}_\perp^2 + z^2}) e^{i q_L z} dz,
\end{equation}
where the longitudinal momentum transfer $q_L$ is determined by the meson mass $M_{J/\psi}$ and rapidity $y$:

\begin{equation}
q_L = \frac{M_{J/\psi} e^y}{2\gamma_c}.
\end{equation}

The total amplitude in coordinate space for a collision with impact parameter $\mathbf{b}$ is constructed by summing the contributions from the two nuclei acting as either photon emitter or target:
\begin{equation}\label{Eq:Ar}
A(y,\mathbf r, \pm \frac{\mathbf b}{2}) = \Gamma_{\gamma A \to J/\psi A}(\mathbf r \pm \frac{\mathbf b}{2}) \times \sqrt{\frac{d^2 N_\gamma (\mathbf r \pm \frac{\mathbf b}{2})}{d^2 r}}.
\end{equation}
Finally, the two-dimensional transverse momentum distribution is obtained via a Fourier transformation:
\begin{equation}\label{Eq:dP}
\frac{d^2 P}{dp_x dp_y} = \frac{1}{2\pi} \left | \int d^2 r \left [A_1 (y , \mathbf r -\frac{\mathbf b}{2}) - A_2 (y , \mathbf r +\frac{\mathbf b}{2}) \right ] e^{i\mathbf p \cdot \mathbf r}  \right |^2.
\end{equation}

The experimental yield also includes contributions from incoherent photo-nucleus interactions involving nuclear dissociation ($A'$). 
The cross section for this process is derived as:
\begin{equation}
\sigma_{\gamma A \to V A'} = \sigma_{\gamma p \to V p} \int d^2\mathbf{x}_{\perp} T(\mathbf{x}_{\perp}) e^{-\frac{1}{2}\sigma_{VN}^{in} T(\mathbf{x}_{\perp})},
\end{equation}
where $T(\mathbf{x}_{\perp})$ is the thickness function of the nucleus, 
$\sigma_{VN}^{in}$ is the inelastic vector meson-nucleon cross section and $B_V$ is 
the slope of the $t$ dependence of the $\gamma + p \to V + p$~\cite{wang2022calculations,klein2017starlight,klein1999exclusive}:
\begin{equation}
\sigma_{VN}^{in} = \sigma_{VN} - \frac{\sigma_{VN}^2}{16\pi B_V}.
\end{equation}

\begin{figure*}[htb]
    \centering

        \begin{minipage}[t]{0.24\textwidth}
        \centering
        \includegraphics[width=\linewidth]{Figures/Zr-beta2-tip_Coherent.pdf}
    \end{minipage}
    \hfill
    \begin{minipage}[t]{0.24\textwidth}
        \centering
        \includegraphics[width=\linewidth]{Figures/Zr-beta2-tip_Total.pdf}
    \end{minipage}
    \hfill 
    \begin{minipage}[t]{0.24\textwidth}
        \centering
        \includegraphics[width=\linewidth]{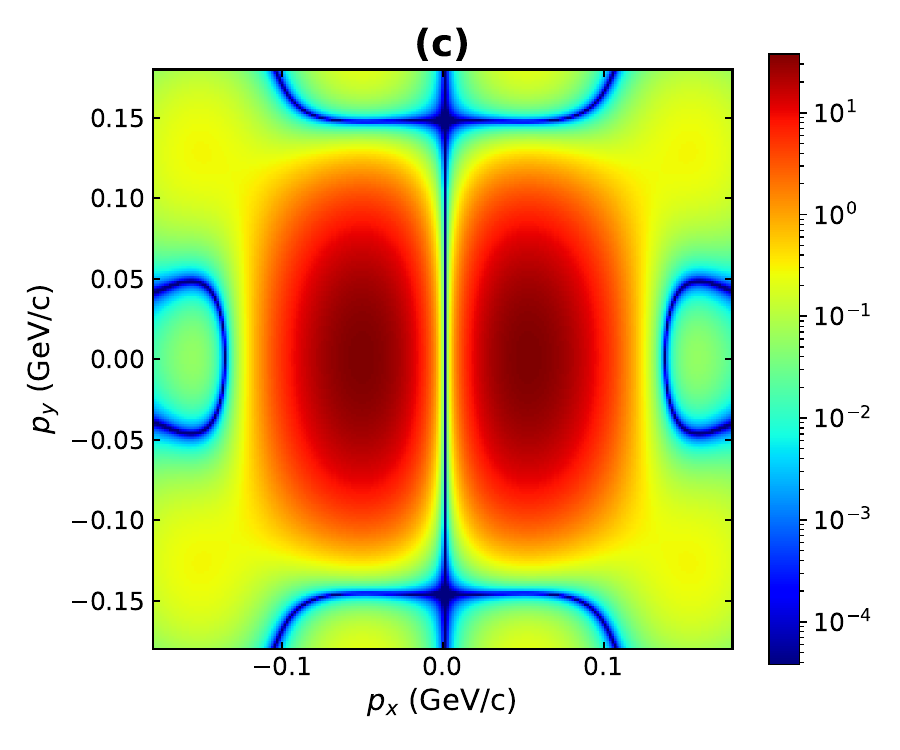}
    \end{minipage}
 \hfill 
    \begin{minipage}[t]{0.24\textwidth}
        \centering
        \includegraphics[width=\linewidth]{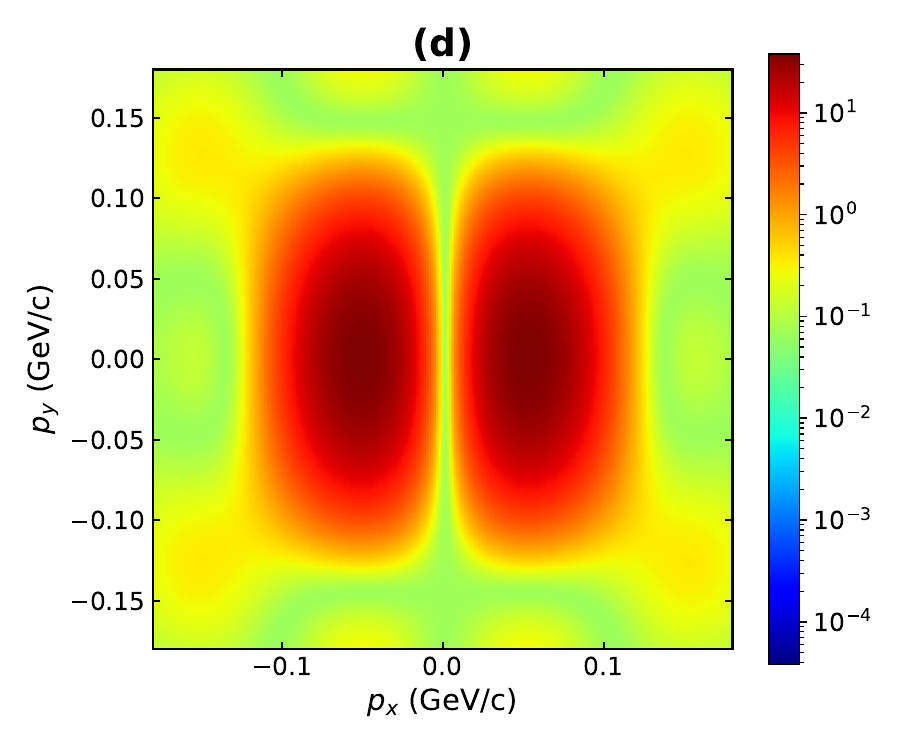}
    \end{minipage}
    \\
    \begin{minipage}[t]{0.24\textwidth}
        \centering
        \includegraphics[width=\linewidth]{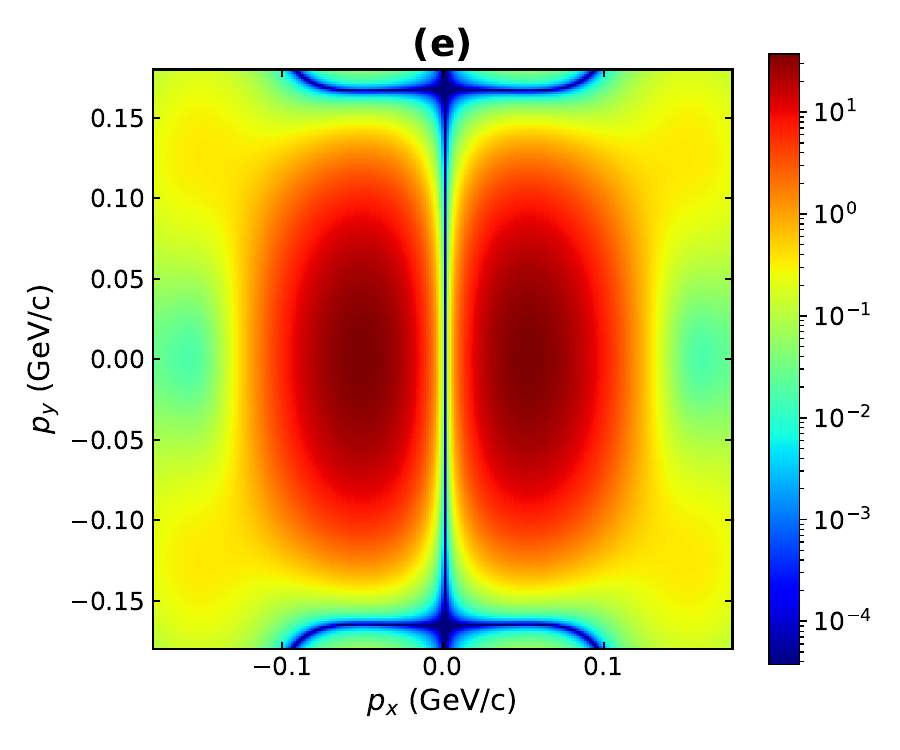}
    \end{minipage}
    \hfill
    \begin{minipage}[t]{0.24\textwidth}
        \centering
        \includegraphics[width=\linewidth]{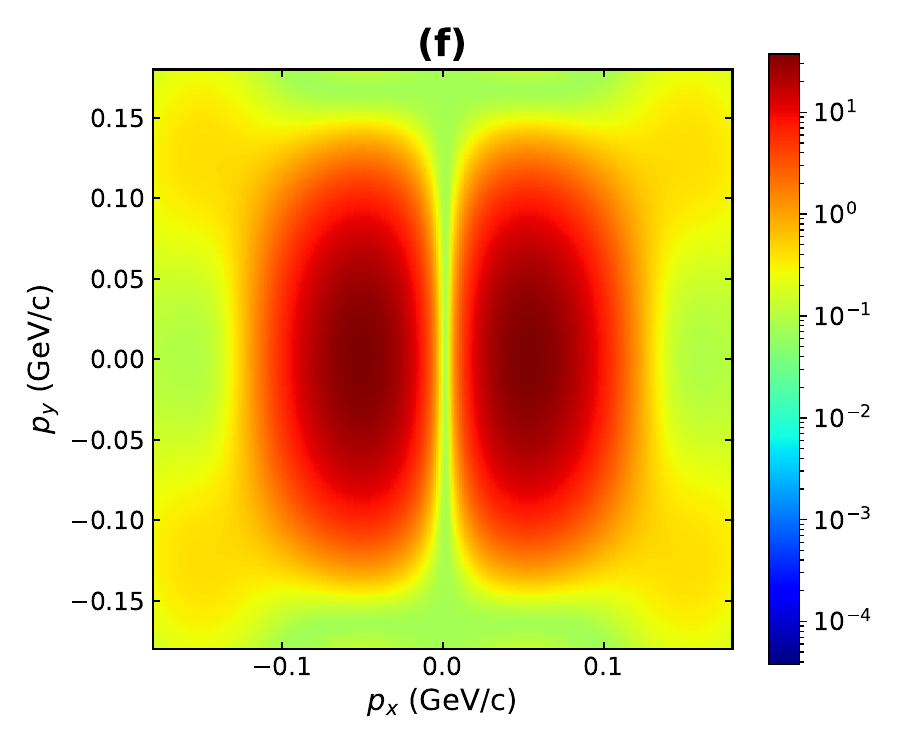}
    \end{minipage}
            \hfill
    \begin{minipage}[t]{0.24\textwidth}
        \centering
        \includegraphics[width=\linewidth]{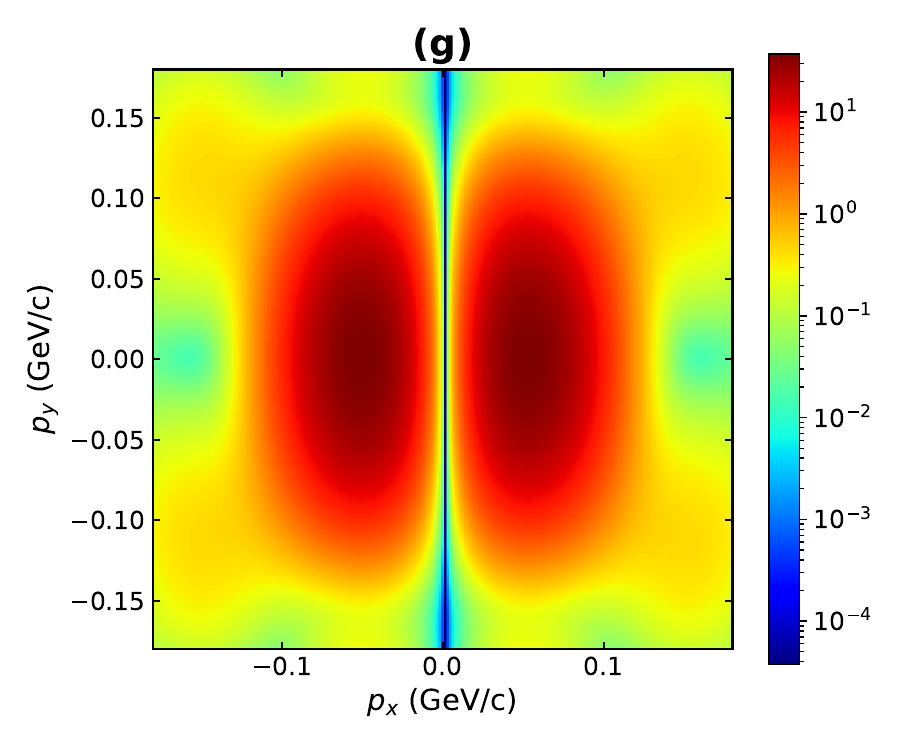}
    \end{minipage}
    \hfill
    \begin{minipage}[t]{0.24\textwidth}
        \centering
        \includegraphics[width=\linewidth]{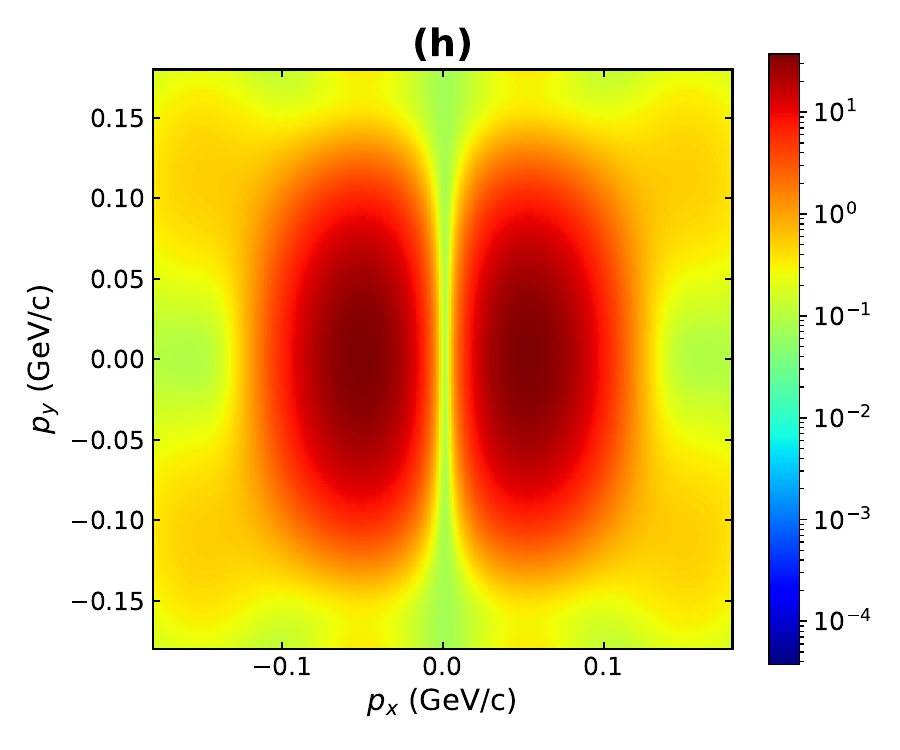}
    \end{minipage}
    \\

    \begin{minipage}[t]{0.24\textwidth}
        \centering
        \includegraphics[width=\linewidth]{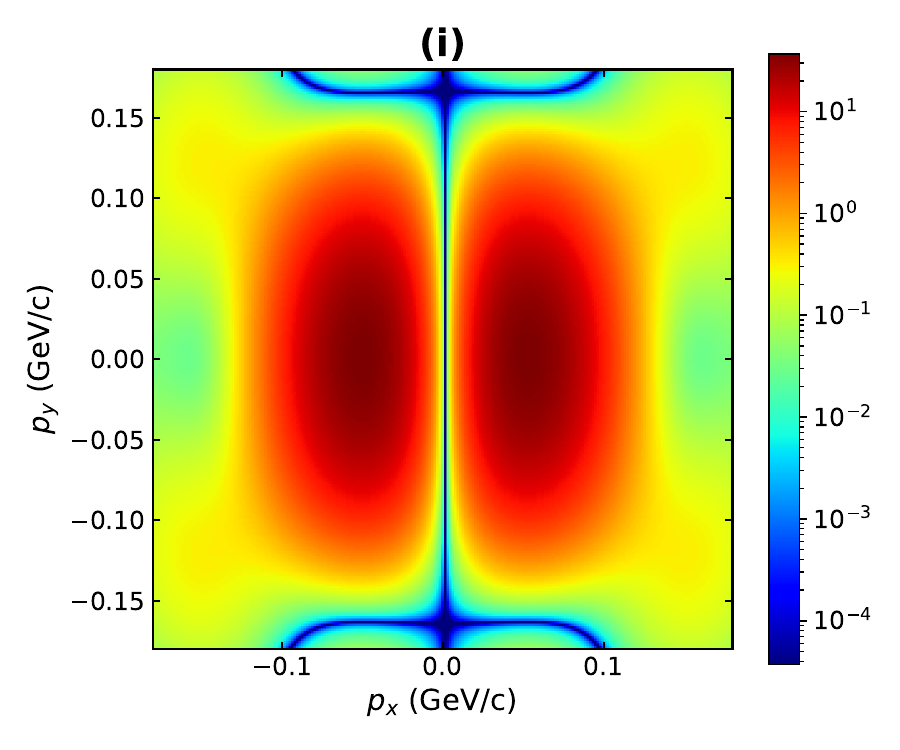}
    \end{minipage}
    \hfill
    \begin{minipage}[t]{0.24\textwidth}
        \centering
        \includegraphics[width=\linewidth]{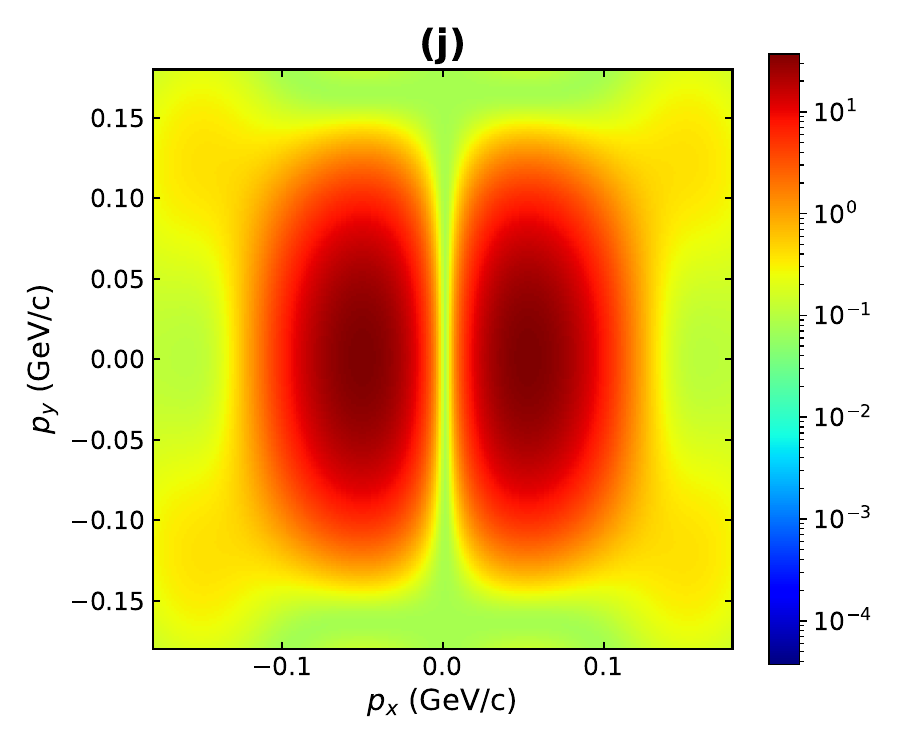}
    \end{minipage}
    \hfill
    \begin{minipage}[t]{0.24\textwidth}
        \centering
        \includegraphics[width=\linewidth]{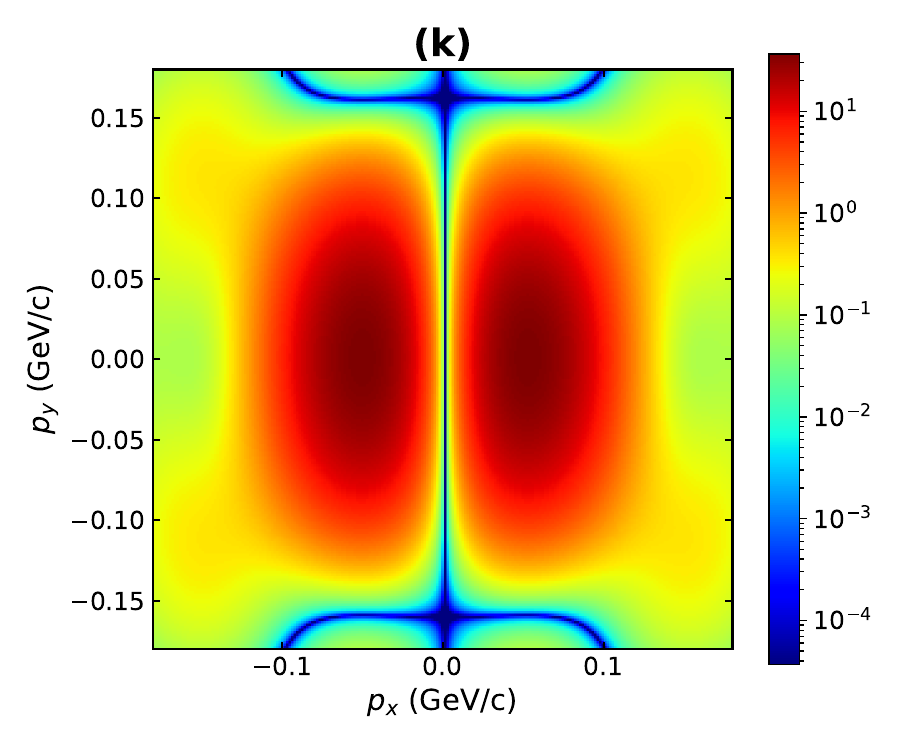}
    \end{minipage}
    \hfill
    \begin{minipage}[t]{0.24\textwidth}
        \centering
        \includegraphics[width=\linewidth]{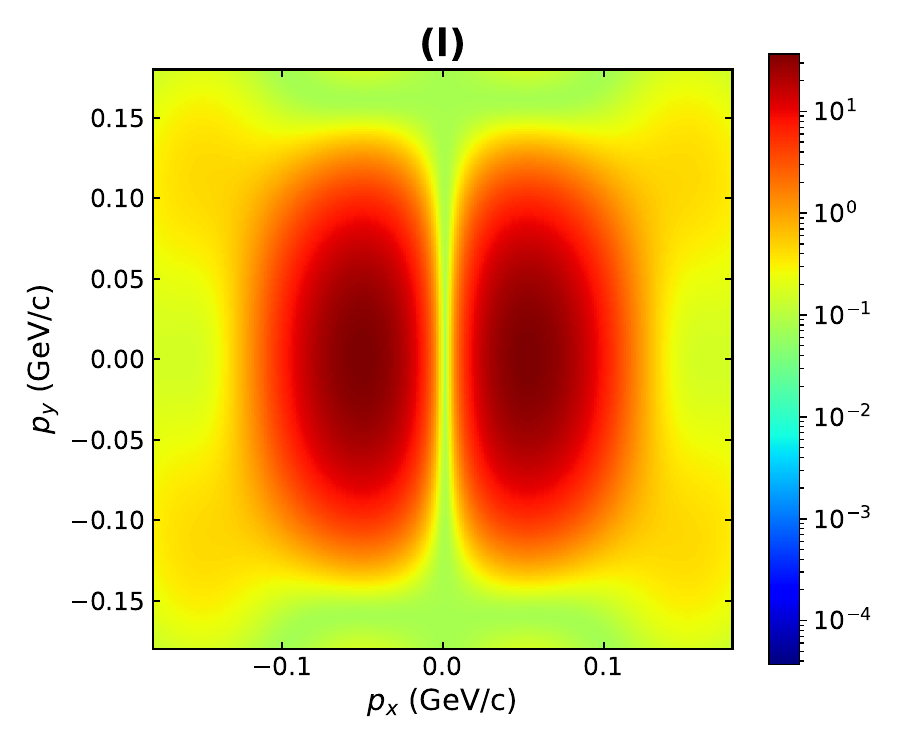}
    \end{minipage}
    \caption{\footnotesize (Color online) 
    Transverse momentum distribution patterns of $J/\psi$ photoproduction in Zr+Zr collisions at mid-rapidity ($y=0$) with an impact parameter of $b = 12$~fm. The panels are organized in pairs corresponding to different nuclear deformation and neutron skin configurations: (a, b) $\beta_2 = 0.3$ tip-tip; (c, d) $\beta_2 = 0.3$ body-body; (e, f) $\beta_3 = 0.2$ tip-tip; (g, h) $\beta_3 = 0.2$ body-body; (i, j) skin-type neutron skin~\cite{xu2021determine}; and (k, l) halo-type neutron skin~\cite{xu2021determine}. In each pair, the left panel represents the coherent process, while the right panel shows the total (coherent + incoherent) result.}
    \label{fig:12_panels_ap}
\end{figure*}

As a complete version of Fig.~\ref{fig:12_panels} in the main text, Fig.~\ref{fig:12_panels_ap} shows the transverse momentum distributions for the full set of deformation and neutron-skin configurations considered in this work.

\section{ SENSITIVITY ANALYSIS OF DEFORMATION SIGNATURES IN MOMENTUM SPACE}\label{sec:deformation_ratio}

\begin{figure*}[htb]
\includegraphics[width=\textwidth]{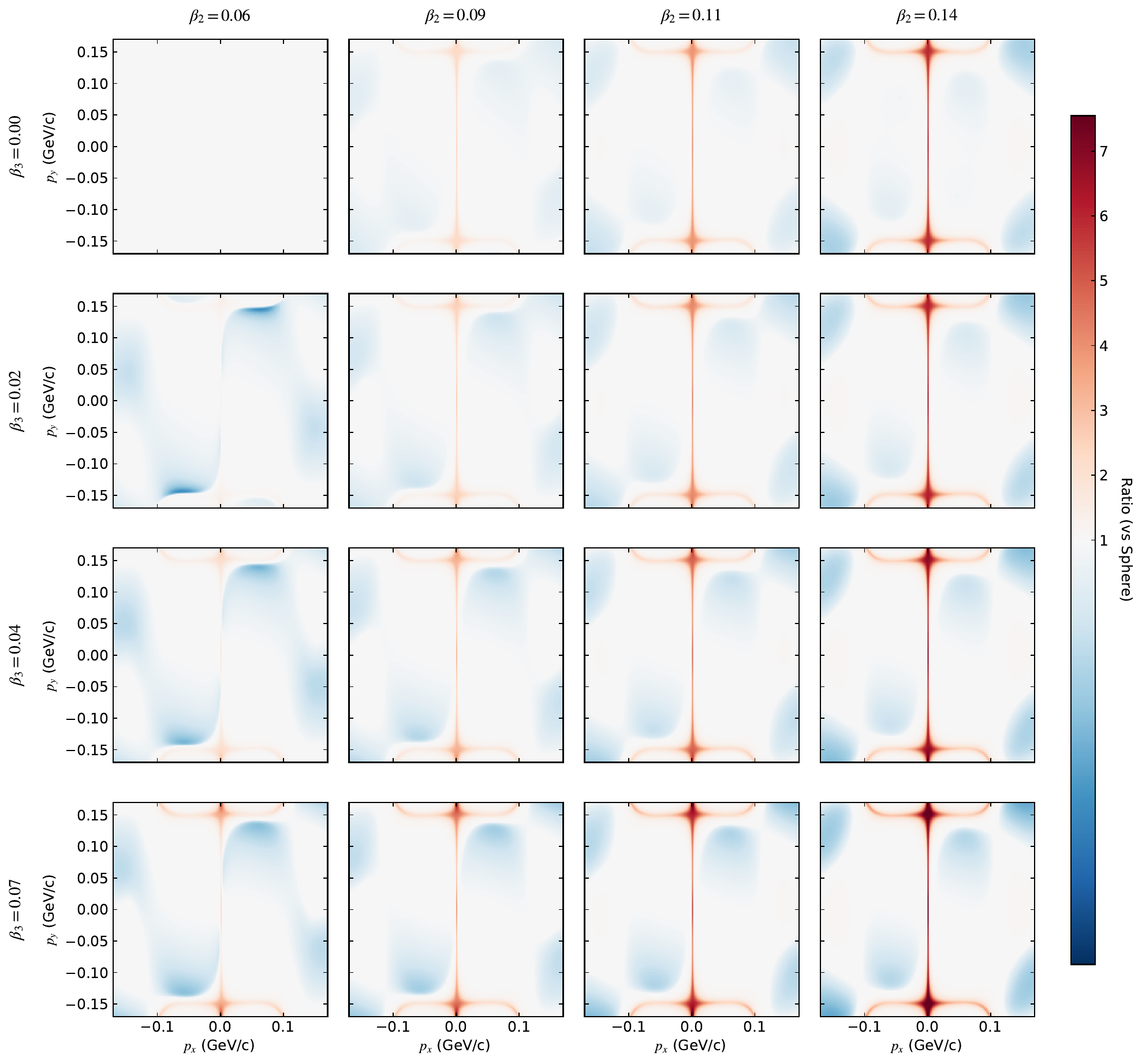}
\caption{\footnotesize(Color online) 
Ratios of the average transverse momentum distributions for deformed nuclei relative to a spherical baseline. This representative $4\times4$ subset illustrates the evolution of the interference pattern with increasing $\beta_2$ and increasing octupole deformation $\beta_3$.}\label{fig:ratio_defor_grid}
\end{figure*}

 Figure~\ref{fig:ratio_defor_grid} shows the ratio of the average $J/\psi$ transverse momentum distributions with nuclear deformation to that of a spherical nucleus. The horizontal axis corresponds to increasing quadrupole deformation ($\beta_2$), while the vertical axis corresponds to increasing octupole deformation ($\beta_3$).
 Increasing $\beta_2$ produces a strong, global enhancement of the central cross-shaped interference dark fringe, whereas increasing $\beta_3$ produces more localized variations that mainly affect the horizontal interference branches. This difference in strength and spatial extent explains the greater resilience of the $\beta_2$ prediction to incoherent background noise compared with $\beta_3$, as seen in Table~\ref{tab:noice_performance} of the main text.

\end{document}